\documentclass{aa}
\usepackage{graphicx}
\usepackage{txfonts}
\usepackage{xcolor}
\usepackage{xspace}
\usepackage[colorlinks=true, citecolor=blue, linkcolor=blue]{hyperref}
\usepackage{perpage} 
\MakePerPage{footnote}
\usepackage{arydshln}
\usepackage{ulem}
\usepackage[switch]{lineno} 
\newcommand{\courrier}[1]{{\fontfamily{lmtt}\selectfont{#1}}}
\newcommand{\ita}[1]{{\textit{#1}}}

\newcommand{\fermi}{\ita{Fermi}-LAT\xspace}

\newcommand{\dg}{$^{\circ}$\xspace}

\newcommand{\gui}[1]{{``{#1}''}}
\newcommand{\hii }{H\,{\sc ii}\xspace}
\newcommand{\hi }{H\,{\sc i}\xspace}
\newcommand{\src }{G150.3+4.5\xspace}
\newcommand{\gam }{$\gamma$-ray\xspace}
\newcommand{\srcbis }{4FGL J0426.5+5434\xspace}

\titlerunning{$\gamma$-ray emission from the SNR \src}
\authorrunning{J. Devin et al.}

\begin{document} 

   \title{High-energy gamma-ray study of the \\ dynamically young SNR G150.3+4.5}

   \author{J. Devin\inst{1}*,
            M. Lemoine-Goumard\inst{1},
            M.-H. Grondin\inst{1},
            D. Castro\inst{2},
            J. Ballet\inst{3},
            J. Cohen\inst{2}
            \and J. W. Hewitt\inst{4}
          }

        \institute{Univ. Bordeaux, CNRS, CENBG, UMR 5797, F-33170 Gradignan, France\\
              *\email{jdevin.phys@gmail.com}
             \and
             Harvard-Smithsonian Center for Astrophysics, 60 Garden Street, Cambridge, MA 02138, USA
             \and
             Laboratoire AIM, CEA-IRFU/CNRS/Universit\'{e} Paris Diderot, D\'{e}partement d’Astrophysique, CEA Saclay, 91191 Gif sur Yvette, France
             \and
             University of North Florida, Department of Physics, 1 UNF Drive, Jacksonville, FL 32224, USA
             }

   \date{Received 26 May 2020, Accepted 11 September 2020}

  \abstract
  {}
  % aims heading (mandatory)
   {The supernova remnant (SNR) G150.3+4.5 was recently discovered in the radio band; it exhibits a shell-like morphology with an angular size of $\sim$ 3\dg, suggesting either an old or a nearby SNR. Extended $\gamma$-ray emission spatially coincident with the SNR was reported in the \ita{Fermi} Galactic Extended Source Catalog, with a power-law spectral index of $\Gamma$ = 1.91 $\pm$ 0.09. Studying particle acceleration in SNRs through their \gam emission is of primary concern to assess the nature of accelerated particles and the maximum energy they can reach.}
  % methods heading (mandatory)
   {Using more than ten years of \fermi data, we investigate the morphological and spectral properties of the SNR \src from 300 MeV to 3 TeV. We use the latest releases of the \fermi catalog, the instrument response functions and the Galactic and isotropic diffuse emissions. We use ROSAT all-sky survey data to assess any thermal and nonthermal X-ray emission, and we derive minimum and maximum distance to \src.}
  % results heading (mandatory)
   {We describe the \gam emission of \src by an extended component which is found to be spatially coincident with the radio SNR. The spectrum is hard and the detection of photons up to hundreds of GeV points towards an emission from a dynamically young SNR. The lack of X-ray emission gives a tight constraint on the ambient density $n_0 \leq 3.6 \times 10^{-3}$ cm$^{-3}$. Since \src is not reported as a historical SNR, we impose a lower limit on its age of $t$ = 1 kyr. We estimate its distance to be between 0.7 and 4.5 kpc. We find that \src is spectrally similar to other dynamically young and shell-type SNRs, such as RX J1713.7$-$3946 or Vela Junior. The broadband nonthermal emission is explained with a leptonic scenario, implying a downstream magnetic field of $B = 5$ $\mu$G and acceleration of particles up to few TeV energies.
   }
   {}


   \maketitle
%
%________________________________________________________________

\section{Introduction}
Supernova remnants (SNRs) have long been thought to be the most likely accelerators of cosmic rays (CRs) up to the knee ($\sim$ 3 $\times$ 10$^{15}$ eV) of the CR spectrum, with diffusive shock acceleration \citep[DSA,][]{Bell:1978} being the primary mechanism accelerating the charged particles to $\gamma$-ray emitting energies. The Large Area Telescope (LAT) on board the \ita{Fermi} satellite has already demonstrated that CR protons can indeed be accelerated at SNR shocks, through detection of the characteristic ``pion bump'' feature due to accelerated protons colliding with ambient matter. However, evidence of CR protons is often found in some evolved SNRs interacting with molecular clouds (MCs), like IC 443 or W44 \citep{Ackermann:2013}, and the maximum particle energy only reaches hundreds of GeV. Cassiopeia A, one of the youngest Galactic SNR ($t \sim$ 350 years) exhibits a \gam spectrum with a low-energy break \citep{Yuan:2013}, characteristic of emission from CR protons, and a photon cutoff energy at a few TeV \citep{CasA:2017, Abeysekara_CasA:2020}. Thus, the question of whether SNRs can accelerate particles up to PeV energies is still open. Moreover, the spectrum of the TeV shell-like and young SNRs can often be accurately described by a leptonic scenario due to inverse Compton (IC) scattering of accelerated electrons on ambient photon fields \citep{Acero:2015}. 

\begin{figure}[t!]
    \centering
    \includegraphics[scale=0.7]{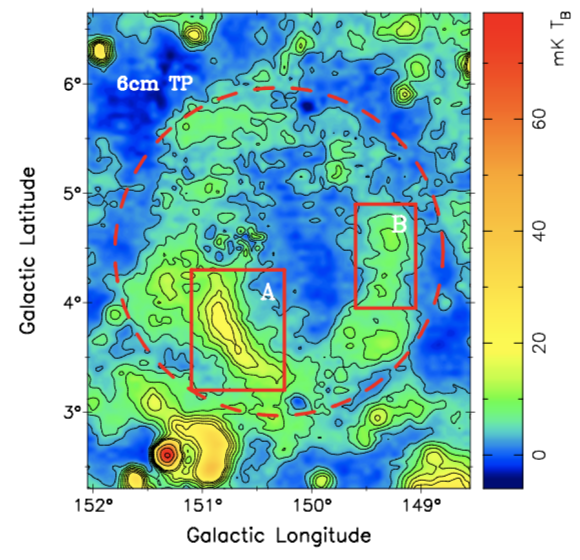}
    \caption{Urumqi 6 cm observations of the SNR \src. The red boxes correspond to the brightest parts of the shell, which is  represented by the red dashed circle. The figure is  from \cite{Gao:2014}.}
    \label{fig:Radiomap}
\end{figure}

G150.3+4.5 was first considered  an SNR candidate, due to the radio detection of the brightest part of the shell, before being firmly identified as a radio SNR. Faint radio emission from the southeastern part of the shell of \src (called G150.8+3.8) was first reported in \cite{Gerbrandt:2014} using the Canadian Galactic Plane Survey (CGPS) and considered a strong SNR candidate due to its semi-circular shell-like appearance and its nonthermal spectrum ($\alpha = -0.62 \pm 0.07$ where $S_{\nu} \propto \nu^{\alpha}$). Using a larger extraction region (128.2$'$ $\times$ 37.6$'$), they derived a radio spectral index of $\alpha = -0.38 \pm 0.10$, centered on $l=150.78$\dg and $b=3.75$\dg. \cite{Gerbrandt:2014} also detected faint red optical filaments spatially coincident with the southeastern shell. \cite{Gao:2014} performed follow-up observations of the region using Urumqi 6 cm survey data (as well as Effelsberg 11 cm and 21 cm data and CGPS 1420 MHz and 408 MHz observations), taking advantage of the survey's extended Galactic latitude range, up to $b=20$\dg. They reported a clear detection of a 2.5\dg wide by 3\dg high ($l=150.3$\dg, $b=+4.5$\dg, as illustrated in Figure~\ref{fig:Radiomap}), synchrotron emitting, shell-like object (named \src), bolstering an SNR origin for the radio emission. The radio spectral indices of the two brightest parts of the shell (southeastern and western sides, red boxes in Figure~\ref{fig:Radiomap}) are $\alpha = -0.40 \pm 0.17$ and $\alpha = -0.69 \pm 0.24$ compatible with the results obtained in \cite{Gerbrandt:2014}.

In the Second Catalog of Hard \fermi Sources \citep[2FHL,][]{2FHL}, an extended source located near the radio SNR was reported (2FHL J0431.2+5553e) and modeled as a uniform disk with a radius $r$ = 1.27\dg $\pm$ 0.04\dg, exhibiting a hard power-law spectral index ($\Gamma$ = 1.7 $\pm$ 0.2) from 50 GeV to 2 TeV. The \ita{Fermi} Galactic Extended Source catalog \citep[FGES,][]{FGES:2017} reported a uniform disk ($r$ = 1.52\dg $\pm$ 0.03\dg), spatially coincident with the radio shape, but with a softer power-law spectrum ($\Gamma$ = 1.91 $\pm$ 0.09 from 10 GeV to 2 TeV) than previously derived. This FGES source is contained in the latest \fermi catalog \citep[4FGL,][]{4FGL:2019}, called 4FGL J0427.2+5533e, and its emission is described by a logarithmic parabola (LP) spectrum. 

In this paper we present a detailed study of the \gam emission in the direction of \src from 300 MeV to 3 TeV. In Section~\ref{sec:fermi} we investigate the morphology of the \gam emission towards \src from 1 GeV to 3 TeV and we explore its spectral properties from 300 MeV to 3 TeV. In Section~\ref{sec:mwl} we use ROSAT observations to assess any thermal and nonthermal X-ray emission from the SNR, and we derive two bounding distance estimates for \src. Finally, in Section~\ref{sec:discussion} we discuss potential \gam emission scenarios and model the broadband nonthermal emission from \src to understand the nature of the accelerated particles.

%____________________________________________________________________________________________
\section{\fermi analysis}\label{sec:fermi}
%____________________________________________________________________________________________
\subsection{Data reduction and preparation}

The \fermi is a pair-conversion instrument that detects \gam photons in an energy range from 20 MeV to higher than hundreds of GeV \citep[a detailed description of the instrument can be found in][]{LAT:Atwood:2009}. We used $\sim$ 10.5 years of \fermi data (from August 4, 2008, to January 31, 2019) and we performed a binned likelihood analysis within a region of interest (ROI) of 23\dg $\times$ 23\dg centered on the SNR \src. A maximum zenith angle of 100\dg was applied to reduce the contamination of the Earth limb, and the time intervals during which the satellite passed through the South Atlantic Anomaly were excluded. Since the point spread function (PSF) of the \fermi is energy dependent and broad at low energy, we started the morphological analysis at 1 GeV while the spectral analysis was made from 300 MeV to 3 TeV. The \courrier{Summedlikelihood} method was used to simultaneously fit events with different angular reconstruction quality (PSF event types). We used version 1.0.10 of the \courrier{Fermitools} and, from 1 GeV to 3 TeV, we used the \courrier{Fermipy} package \citep[version 0.17.4,][]{Wood:2017} that allows for simultaneous morphological and spectral fits, setting a pixel size to 0.1\dg and eight energy bins per decade. The spectral analysis was made from 300 MeV to 3 TeV with ten energy bins per decade and a pixel size of 0.1\dg. For the entire analysis, we used the \courrier{SOURCE} event class and a preliminary inflight corrected version of the P8R3\_V2 instrument response functions (IRFs) that corrects for a $\sim$10\% discrepancy of the PSF event type effective areas. For consistency with the 4FGL analysis, the energy dispersion was taken into account except for the isotropic diffuse emission. 

To model the \gam data around \src, we started with the 4FGL catalog keeping in the model the sources located within a 20\dg radius from the ROI center. We used the \courrier{gll\_iem\_v07.fits} model to describe the Galactic diffuse emission and the isotropic templates derived using the preliminary inflight corrected version of the P8R3\_V2 IRFs and extrapolated up to 3 TeV.\footnote{\fermi background models can be found at \url{https://fermi.gsfc.nasa.gov/ssc/data/access/lat/BackgroundModels.html}.} The spectral parameters of the sources in the model were first fit simultaneously with the Galactic and isotropic diffuse emissions from 1 GeV to 3 TeV. To search for additional sources in the ROI, we computed a residual test statistic (TS) map that tests in each pixel the significance of a source with a generic $E^{-2}$ spectrum against the null hypothesis:
\begin{equation}
    \rm{TS} = 2 \times (\rm{log} \mathcal{L_{\rm{1}}} - \rm{log} \mathcal{L_{\rm{0}}})
.\end{equation}
Here $\mathcal{L_{\rm{1}}}$ and $\mathcal{L_{\rm{0}}}$ are the likelihood obtained with the model including and excluding the source, respectively. The TS follows a $\chi^2$ distribution with $n$ degrees of freedom corresponding to the additional free parameters between the models 1 and 0. We iteratively added three point sources in the model where the TS exceeded 25. We localized the position of these additional sources (RA$-$Dec = $52.08^{\circ}-53.51^{\circ}$, $76.62^{\circ}-45.76^{\circ}$ and $79.43^{\circ}-50.49^{\circ}$) and we fit their spectral parameters simultaneously with the Galactic and isotropic diffuse emissions. The residual TS map obtained with all the sources considered in the model shows no significant residual emission, indicating that the ROI is adequately modeled.

\begin{table*}[t!]
    \centering
    \begin{tabular}{lcccccccc}
        \hline
        \hline
               & RA$_{\rm{J2000}}$ (\dg)  & Dec$_{\rm{J2000}}$ (\dg)   & $\sigma$ or $r$ (\dg)  &     $r_{68}$ (\dg) & TS     & TS$_{\rm{ext}}$ &  N$_{\rm{dof}}$ & \\
        \hline
            Gaussian  & 66.425 $\pm$   0.060     &  55.371 $\pm$   0.056    
             &  0.900$_{-0.028}^{+0.029}$        & 1.359$_{-0.043}^{+0.044}$ 
             &  624.2     & 549.8 &   6 \\
             Disk      &   66.555 $\pm$ 0.172   &  55.314 $\pm$   0.150    
             & 1.497$_{-0.028}^{+0.032}$     &  1.235$_{-0.023}^{+0.026}$
             &  595.0      &   520.7  &  6  \\
             Urumqi 6 cm map   & -- & -- & -- & -- & 361.7 & 287.3 & 3 \\
        \hline
    \end{tabular}
    \caption{Best-fit positions and extensions (1 GeV $-$ 3 TeV) of \src ($r_{68}$ being the 68\% containment radius) with the associated statistical errors, using a 2D symmetric Gaussian and a uniform disk model. The TS values and the number of degrees of freedom (N$_{\rm{dof}}$) are given with respect to the null hypothesis (no emission from \src). The result using the Urumqi 6 cm map of the SNR, represented by the yellow contours in Figure~\ref{fig:morpho}, is also given.}
    \label{tab:morpho}
\end{table*}

\begin{figure*}[t!]
    \includegraphics[scale=0.45]{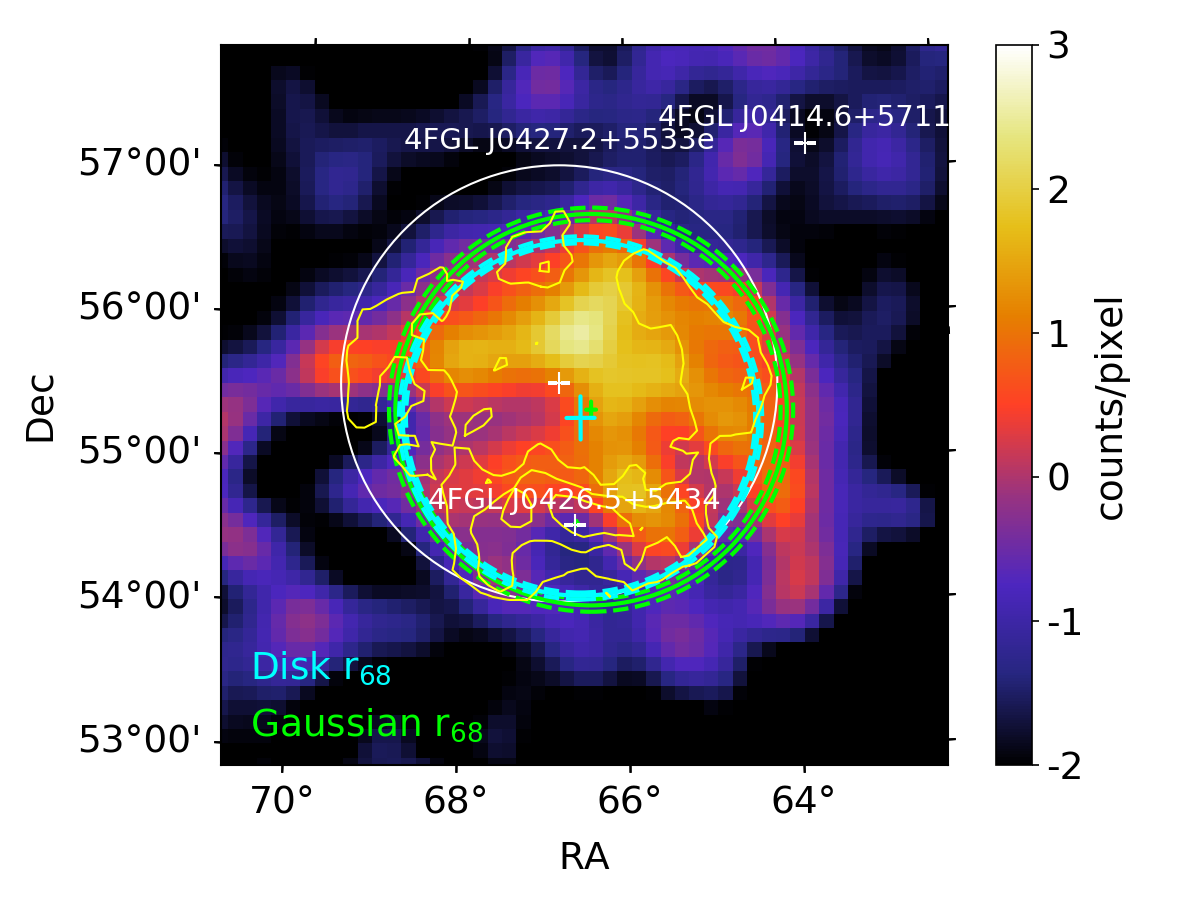}
    \includegraphics[scale=0.45]{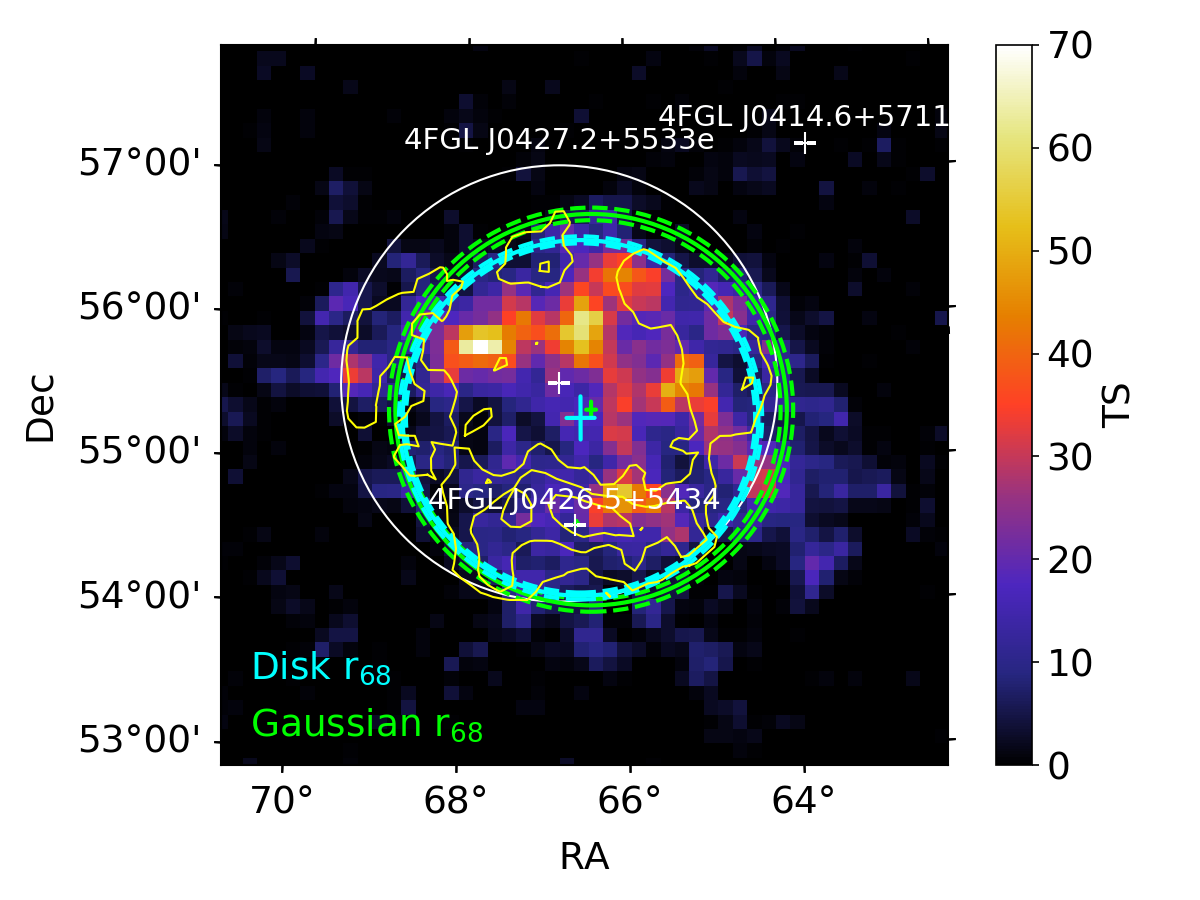}
    \caption{Residual count map (left, smoothed with a Gaussian kernel of 0.2\dg) and residual TS map (right) obtained from 1 GeV to 3 TeV without \src included in the model. The best-fit Gaussian and disk are represented in green and blue, respectively. The crosses are the centroid uncertainties (1$\sigma$), while the solid and dashed circles correspond to the $r_{\rm{68}}$ with its associated statistical errors (1$\sigma$). The 4FGL sources are shown in white and the Urumqi 6 cm radio contours (at 6, 11, and 16 mK T$_{\rm{B}}$) are overlaid in yellow.}
    \label{fig:morpho}
\end{figure*}

\subsection{Morphological analysis}\label{sec:fermi_morpho}

In the 4FGL catalog, a point source (\srcbis) is located in the southern part of the SNR with a significance of $\sim$ 25$\sigma$ between 100 MeV and 1 TeV and a LP spectrum. We performed the morphological analysis from 1 GeV to 3 TeV, with the free parameters in the model being the normalization of the sources located closer than 5\dg from the ROI center, of the Galactic and isotropic diffuse emissions and the spectral parameters of \src and \srcbis.

Using the morphological parameters of \src reported in the 4FGL catalog, we first localized the position of \srcbis and we obtained RA$_{\rm{J2000}}$ = 66.61\dg $\pm$ 0.02\dg and Dec$_{\rm{J2000}}$ = 54.58\dg $\pm$ 0.02\dg which is similar to the values reported in the 4FGL catalog (RA$_{\rm{J2000}}$ = 66.63\dg and Dec$_{\rm{J2000}}$ = 54.57\dg). We tested its extension by localizing a 2D symmetric Gaussian. The significance of the extension is calculated through TS$_{\rm{ext}}$ = 2 $\times$ (log$\mathcal{L_{\rm{ext}}}$ $-$ log$\mathcal{L_{\rm{PS}}}$) where $\mathcal{L_{\rm{ext}}}$ and $\mathcal{L_{\rm{PS}}}$ are the likelihood obtained with the extended and point-source model, respectively. Since the extended model adds only one parameter to the point-source model (the sigma extent), the significance of the extension is $\sqrt{\rm{TS_{ext}}}$. A source is usually considered significantly extended if TS$_{\rm{ext}} \geq 16$ (corresponding to 4$\sigma$). For \srcbis, we found TS$_{\rm{ext}} = 7.7$ indicating that the emission is not significantly extended. 

\begin{figure*}[th!]
    \includegraphics[scale=0.45]{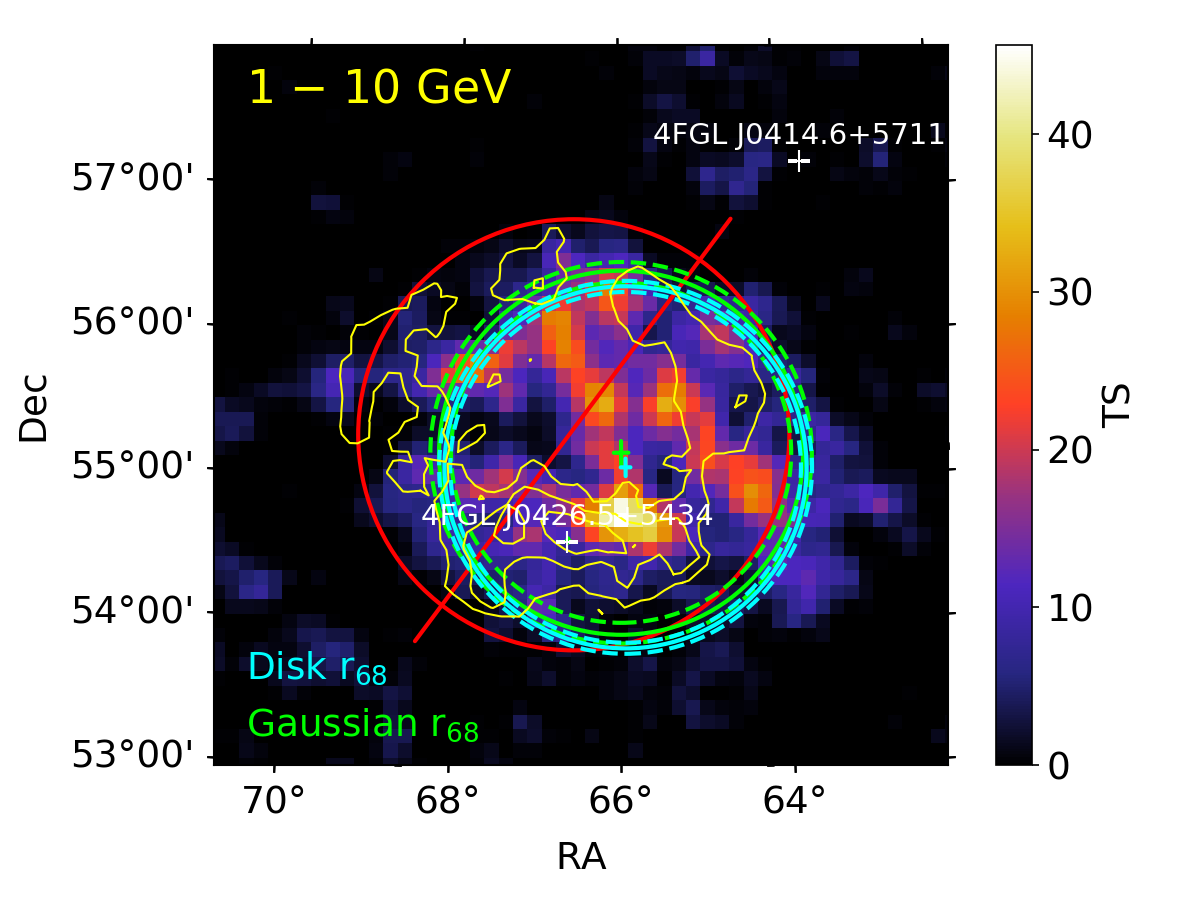}
    \includegraphics[scale=0.45]{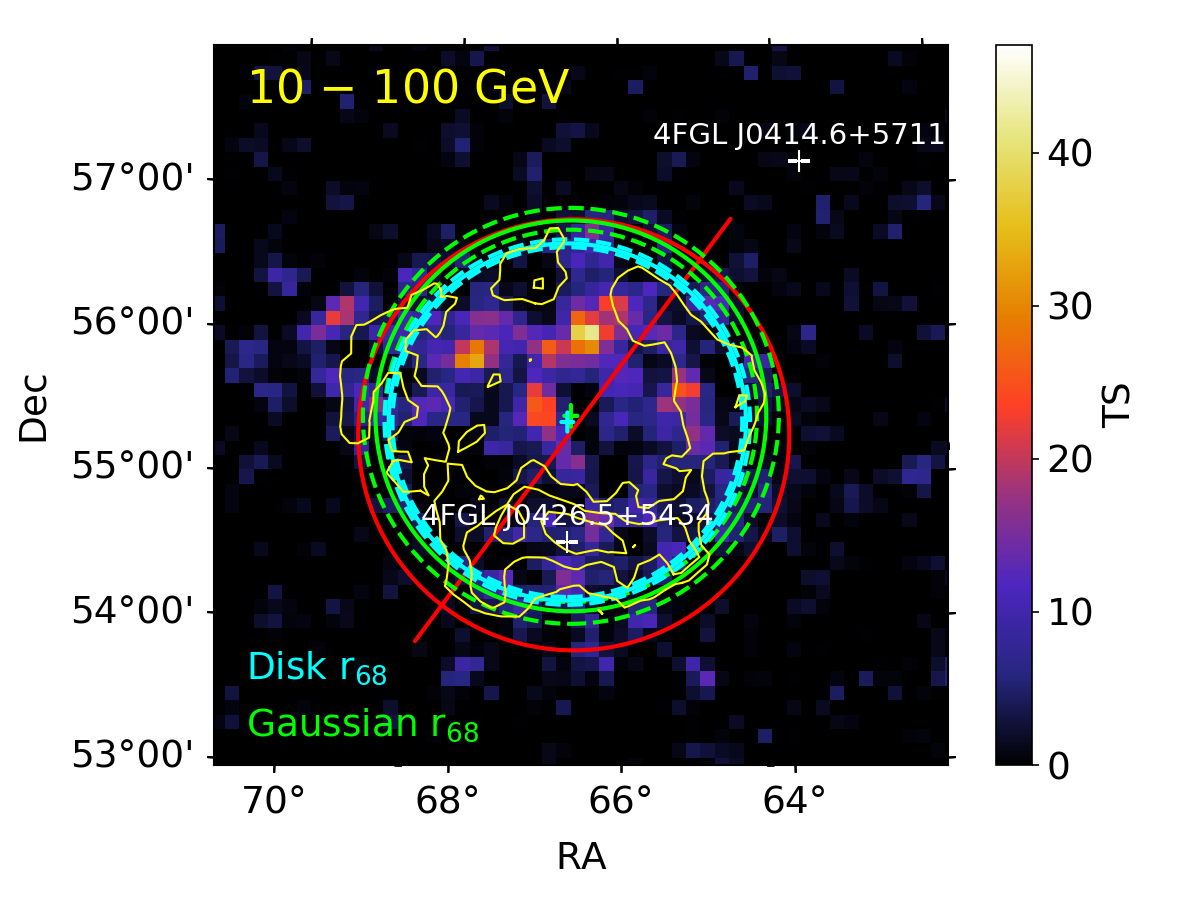}
    \caption{Residual TS maps without \src in the model from 1 to 10 GeV (left panel) and 10 to 100 GeV (right panel). The best-fit Gaussian and disk are represented in green and blue, respectively, the crosses being the centroid uncertainties (1$\sigma$) and the solid and dashed circles corresponding to the $r_{\rm{68}}$ with its associated statistical errors (1$\sigma$). The white symbols and yellow contours are the same as in Figure~\ref{fig:morpho}. The red circle and line show the northeastern and southwestern hemispheres of the best-fit disk found between 1 GeV and 3 TeV (see Section~\ref{sec:fermi_spec}).}
    \label{fig:Edpt_morpho}
\end{figure*}

We then tested two spatial models to describe the emission of \src: a uniform disk and a 2D symmetric Gaussian. The best-fit position of \srcbis was initially derived with a disk model for \src, following the procedure in the 4FGL catalog. In order to account for possible spatial variations of this point source when using a different spatial model for \src, we fit iteratively (two times) the morphological parameters of \src (using either the disk or the Gaussian model) and those of \srcbis. The best-fit position of \srcbis using the Gaussian model was similar to the one using the disk (with a likelihood variation of 0.2). So we fixed the position of \srcbis to RA$_{\rm{J2000}}$ = 66.61\dg and Dec$_{\rm{J2000}}$ = 54.58\dg for both spatial models of \src. 

Table~\ref{tab:morpho} reports the best-fit positions and extensions obtained for the 2D symmetric Gaussian and disk models, with the associated TS and TS$_{\rm{ext}}$ values. The Akaike criterion \citep{Akaike} indicates a preference for the  2D symmetric Gaussian model compared to the uniform disk model ($\Delta{\rm{TS_{Disk \rightarrow Gaussian}}} = 29.2$). A possible source of systematic uncertainties is related to the specific Galactic diffuse model used in the analysis. Thus, we repeated the analysis using the former diffuse emission models (\courrier{gll\_iem\_v06.fits} and the associated isotropic component file \courrier{iso\_P8R3\_SOURCE\_V2.txt}). We refit all the sources in the ROI to account for possible spectral differences induced by the different background model before doing the morphological analysis. We still found that the \gam emission is best described by a 2D symmetric Gaussian. 

We also replaced the extended component by the Urumqi 6 cm map of the SNR \citep[Figure~\ref{fig:Radiomap},][]{Gao:2014} and we obtained $\Delta{\rm{TS_{Radio \rightarrow Gaussian}}} = 262.5$ (see Table~\ref{tab:morpho}). The residual TS map after fitting the radio template shows significant emission in the inner part of the shell. We therefore tested a two-component model, involving the radio map of the SNR and a 2D symmetric Gaussian, whose best-fit spatial parameters were found to be similar to those listed in Table~\ref{tab:morpho}. We found $\Delta{\rm{TS_{Gaussian \rightarrow Gaussian+Radio}}}=1.4$ for three additional parameters, indicating that the radio template does not fit the gamma-ray morphology.

Figure~\ref{fig:morpho} depicts the residual count map and TS map without \src included in the model with the best-fit spatial models and the radio contours overlaid. The radio emission is spatially coincident with the \gam emission, although the brightest parts of the radio and \gam emission do not arise from the same region. Residual count maps in different energy bands without \src and \srcbis included in the model (not shown here) indicate that \srcbis is brighter than \src below 3 GeV, while the emission from the SNR is dominant at higher energies ($E > 3$ GeV). 

We also investigated a possible energy-dependent morphology by fitting a Gaussian and a disk model between 1$-$10 GeV and 10$-$100 GeV. The comparison between the two spatial models gives $\Delta{\rm{TS_{Disk \rightarrow Gaussian}}} = -13.7$ (1$-$10 GeV) and $\Delta{\rm{TS_{Disk \rightarrow Gaussian}}} = 19.5$ (10$-$100 GeV). Figure~\ref{fig:Edpt_morpho} shows the residual TS maps without \src in the model between 1$-$10 GeV and 10$-$100 GeV, with the best-fit spatial models overlaid. The low-energy \gam emission ($E < 10$ GeV) is dominated by the southwestern part of the SNR, while the centroid of the best-fit spatial models displaces towards the center of the SNR at higher energies ($E > $ 10 GeV). We note that the best-fit models from 10 GeV to 100 GeV are very similar to those found between 1 GeV and 3 TeV (Figure~\ref{fig:morpho}). The displacement of the centroid of the best-fit models at low energy may be due to a possible contamination from \srcbis. Using the Gaussian model, the centroid moves from RA = 65.98\dg $\pm$ 0.09\dg and Dec = 55.19\dg $\pm$ 0.08\dg (1$-$10 GeV) to RA = 66.59\dg $\pm$ 0.08\dg and Dec = 55.45\dg $\pm$ 0.08\dg (10$-$100 GeV). The \gam morphology does not shrink at higher energies with the Gaussian extents between 1$-$10 GeV ($r_{68} = 1.26$\dg$_{-0.08^{\circ}}^{+0.06^{\circ}}$) and 10$-$100 GeV ($r_{68} = 1.36$\dg$_{-0.06^{\circ}}^{+0.09^{\circ}}$) being compatible within statistical errors.

\begin{table*}[th!]
    \centering
    \tiny{
    \begin{tabular}{lcccccc}
        \hline
        \hline
                       & $N_{0}$ (ph cm$^{-2}$ s$^{-1}$ MeV$^{-1}$) & $\alpha$  & $\beta$  &   $E_{0}$ (MeV) & $\Phi$ & $F$ \\
        \hline
             \src - Gaussian      &  (7.78 $\pm$ 0.46$_{\rm{stat}}$ $\pm$ 1.61$_{\rm{syst}}$) $\times$ 10$^{-14}$  &  1.62 $\pm$ 0.04$_{\rm{stat}}$ $\pm$ 0.22$_{\rm{syst}}$     
                        &  0.07 $\pm$  0.02$_{\rm{stat}}$ $\pm$ 0.02$_{\rm{syst}}$  & 8973.8  & $6.46_{-3.54}^{+10.12}$ & $9.75_{-5.84}^{+36.41}$ \\
             \srcbis    &   (1.24 $\pm$ 0.06$_{\rm{stat}}$ $\pm$ 0.07$_{\rm{syst}}$) $\times$ 10$^{-11}$   &  2.48 $\pm$ 0.08$_{\rm{stat}}$ $\pm$ 0.10$_{\rm{syst}}$    
                        &  0.41 $\pm$ 0.08$_{\rm{stat}}$ $\pm$ 0.11$_{\rm{syst}}$    &  651.2 & $14.92_{-2.62}^{+3.14}$ & $1.52_{-0.28}^{+0.52}$  \\
        \hdashline
        \src - Disk      &  (6.58 $\pm$ 0.40$_{\rm{stat}}$ $\pm$ 1.20$_{\rm{syst}}$) $\times$ 10$^{-14}$  &  1.68 $\pm$ 0.04$_{\rm{stat}}$ $\pm$ 0.21$_{\rm{syst}}$     
                        &  0.06 $\pm$  0.02$_{\rm{stat}}$ $\pm$ 0.03$_{\rm{syst}}$  & 8973.8 & $6.34_{-3.61}^{+11.10}$ & $7.95_{-4.64}^{+33.88}$ \\
             \srcbis    &   (1.22 $\pm$ 0.06$_{\rm{stat}}$ $\pm$ 0.06$_{\rm{syst}}$) $\times$ 10$^{-11}$   &  2.47 $\pm$ 0.08$_{\rm{stat}}$ $\pm$ 0.11$_{\rm{syst}}$    
                        &  0.41 $\pm$ 0.08$_{\rm{stat}}$ $\pm$ 0.12$_{\rm{syst}}$    &  651.2 & $14.70_{-2.53}^{+3.06}$ & $1.50_{-0.28}^{+0.54}$ \\
        \hline
    \end{tabular}}
    \caption{Best-fit spectral parameters (from 300 MeV to 3 TeV) with the associated errors obtained for \src and \srcbis using a logarithmic parabola spectrum, as defined in Equation~\ref{eq:spec}. The first two and last two lines are the parameters obtained when describing the emission from \src with a Gaussian and a disk respectively. $\Phi$ and $F$ are the integral flux and energy flux from 300 MeV to 3 TeV in units of $10^{-9}$ cm$^{-2}$ s$^{-1}$ and $10^{-11}$ erg cm$^{-2}$ s$^{-1}$, respectively.}
    \label{tab:spec}
\end{table*}

\begin{figure*}[th!]
    \includegraphics[scale=0.31]{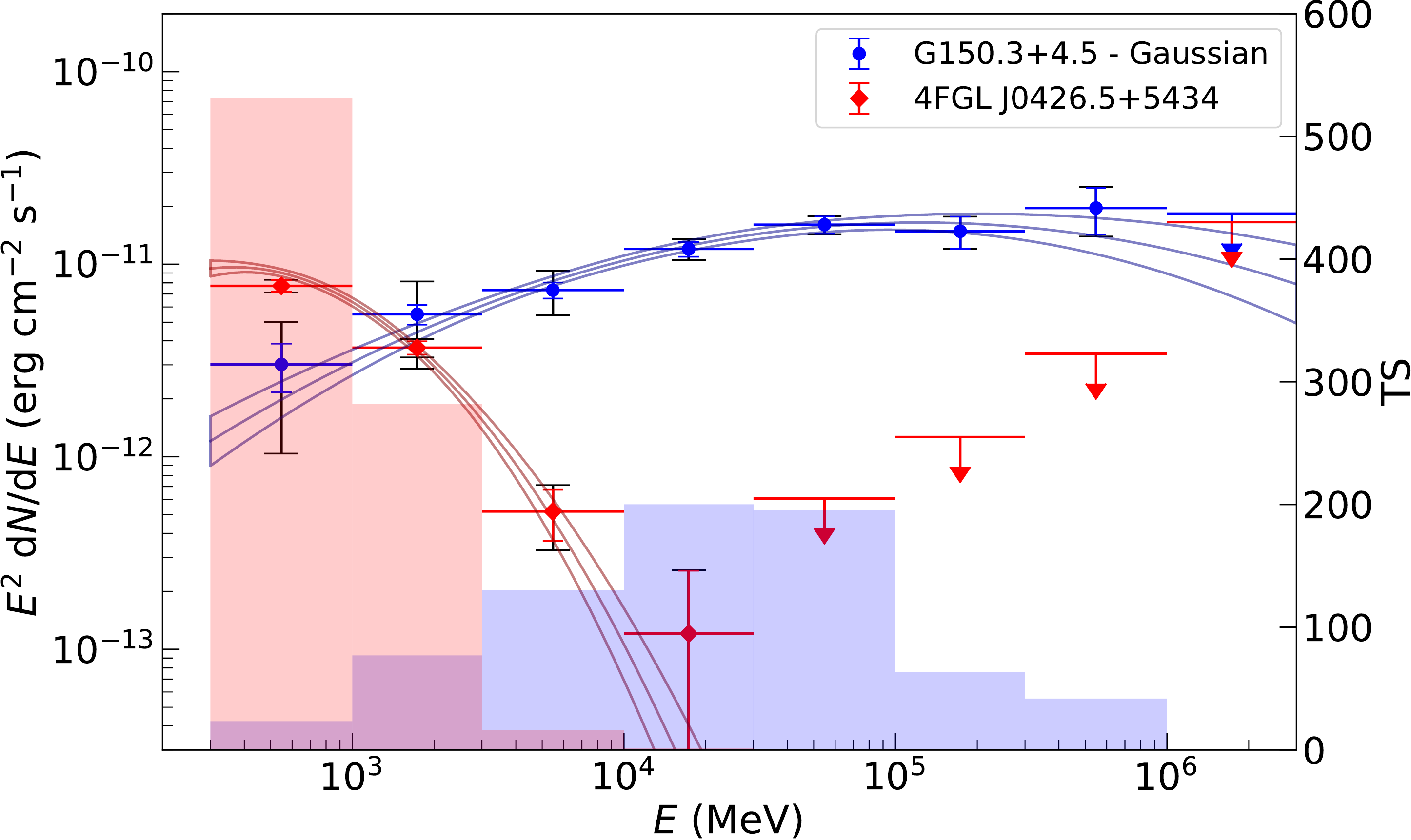}
    \hspace{0.2cm}
    \includegraphics[scale=0.31]{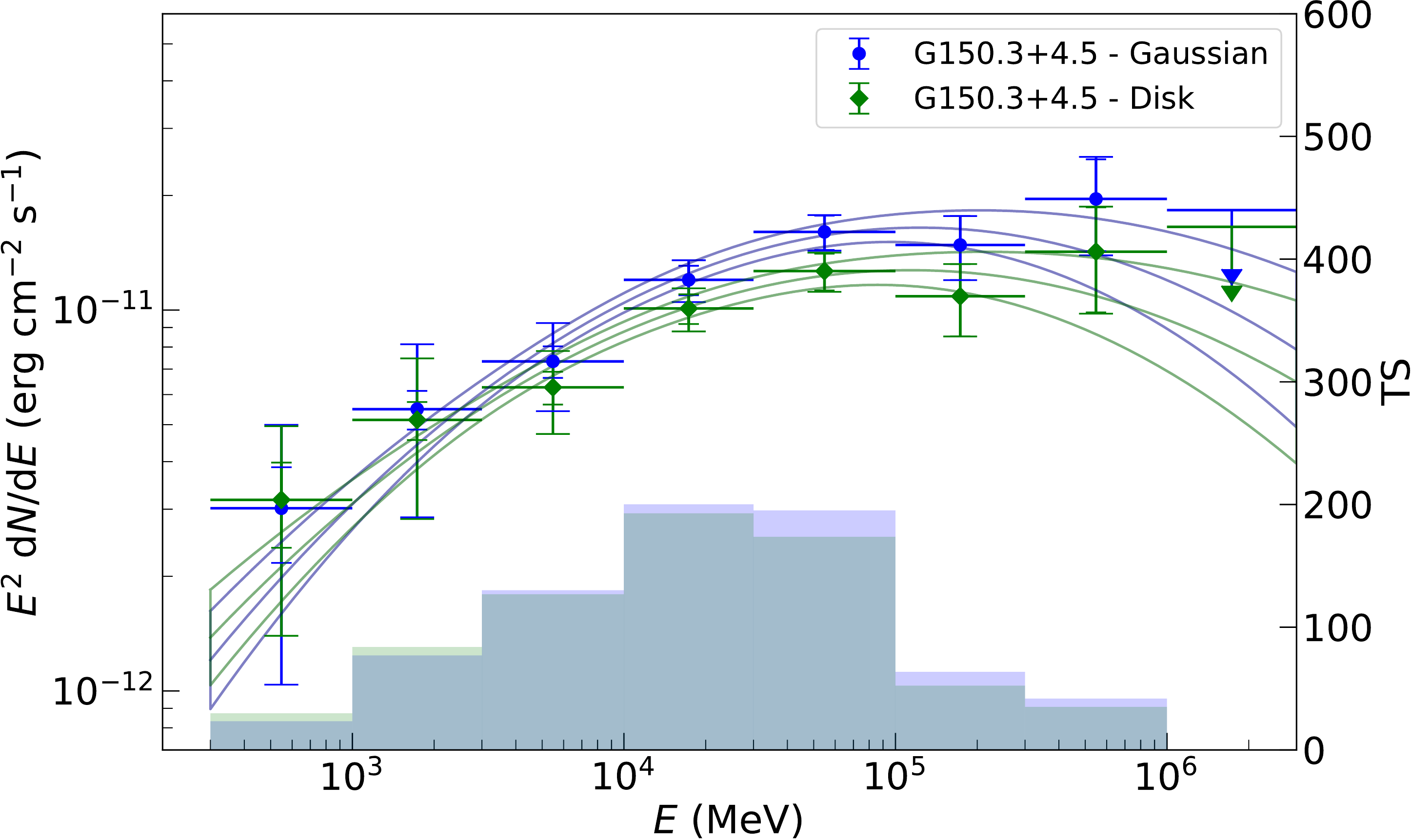}
    \caption{(Left) SEDs of \src (blue circles) and \srcbis (red diamonds) with the associated statistical errors (colored bars) and the quadratic sum of the statistical and systematic errors (black bars). The shaded areas correspond to the TS value in each energy band, and the upper limits are calculated at the 95\% confidence level. (Right) Comparisons of the SEDs obtained using a Gaussian (blue circles) and a disk (green diamonds) model for \src. For visibility purpose, the colored bars represent the quadratic sum of the statistical and systematic errors.}
    \label{fig:SEDs}
\end{figure*}

\subsection{Spectral analysis}\label{sec:fermi_spec}

Using the best-fit spatial model found in Section~\ref{sec:fermi_morpho} (the Gaussian shape for \src), we performed the spectral analysis from 300 MeV to 3 TeV. For comparison, we also performed the same analysis with the disk model. We first fit the spectral parameters of the sources located up to 7\dg from the ROI center and the normalization of all sources in the model, simultaneously with the Galactic and isotropic diffuse emissions. We verified that no additional sources were needed in the model by examining the residual count and TS maps. When testing different spectral shapes for \src and \srcbis, the spectral parameters of sources located up to 5\dg from the ROI center were left free during the fit, like those of the Galactic and isotropic diffuse emissions. In the 4FGL catalog, the emission from \src and \srcbis is described by a LP spectrum of the form

\begin{equation}
\frac{dN}{dE} = N_{0} \times \bigg ( \frac{E}{E_0} \bigg ) ^{- (\alpha + \beta \text{log}(E/E_0) )}
\label{eq:spec}
.\end{equation}

Given our best-fit LP spectral parameters for \srcbis, reminiscent of a pulsar, we first replaced its LP spectrum by a power law with an exponential cutoff that did not significantly improve the fit. Keeping the LP to describe the emission of \srcbis, we then tried a power-law spectrum for \src compared to the LP spectrum. We obtained a significant curvature of the spectrum with $\Delta{\rm{TS_{PL \rightarrow LP}}} = 23.3$  ($\sim$ 4.8$\sigma$). We also tested a broken power-law spectrum that did not significantly improve the fit compared to the LP spectrum. The spectra of \src and \srcbis are thus best described with a LP. We repeated the procedure with the model using the disk component and we obtained the same results. The best-fit spectral parameters are listed in Table~\ref{tab:spec}, with the associated statistical and systematic errors. The systematic errors on the spectral analysis depend on our uncertainties on the Galactic diffuse emission model and on the effective area. The former is calculated using eight alternative diffuse emission models following the same procedure as in the first \fermi supernova remnant catalog \citep{SNR:2016} and the latter is obtained by applying two scaling functions on the effective area\footnote{\url{https://fermi.gsfc.nasa.gov/ssc/data/analysis/scitools/Aeff_Systematics.html}}. We also considered the impact on the spectral parameters when changing the spatial model from Gaussian to disk. All these sources of systematic uncertainties (related to the Galactic diffuse emission model, the effective area, and the spatial model) were thus added in quadrature.

The spectrum of \src is hard, with $\alpha$ = 1.62 $\pm$ 0.04$_{\rm{stat}}$ $\pm$ 0.23$_{\rm{syst}}$ at $E_{0}$ = 9.0 GeV, similar to the value  obtained for young TeV shell-type SNRs such as RX J1713.7$-$3946 or RCW 86 \citep{Acero:2015}. With a soft spectral index $\alpha$ = 2.48 $\pm$ 0.08$_{\rm{stat}}$ $\pm$ 0.10$_{\rm{syst}}$ at $E_{0}$ = 651.2 MeV and a large curvature, \srcbis has a spectrum reminiscent of a pulsar. \cite{Barr:2013} found no pulsations from this source with the 100-m Effelsberg radio telescope, making the nature of \srcbis unclear.

Since the centroid of the best-fit spatial models displaces with energy (Section~\ref{sec:fermi_morpho}), we investigated possible spectral differences between the southwestern (SW) and northeastern (NE) parts of the emission, dividing the best-fit disk (found between 1 GeV and 3 TeV) into two components (red circle and line in Figure~\ref{fig:Edpt_morpho}). We found a softer spectrum in the SW part than in the NE part, with $\alpha_{\rm{NE}} = 1.57 \pm 0.07$ ($\beta_{\rm{NE}} = 0.08 \pm 0.03$) and $\alpha_{\rm{SW}} = 1.82 \pm 0.06$ ($\beta_{\rm{SW}} = 0.04 \pm 0.03$), confirming the morphological analysis in the 1$-$10 GeV and 10$-$100 GeV bands. We therefore obtained spectral differences between the NE and SW parts, although both regions exhibit a hard spectrum.

We then computed the spectral energy distributions (SEDs) of the Gaussian and the disk components dividing the whole energy range (300 MeV $-$ 3 TeV) into eight bins. We described the emission of \src and \srcbis by a power law with fixed spectral index ($\Gamma$ = 2) to avoid any dependence on the spectral models. The normalizations of \src and \srcbis were fit simultaneously with the Galactic and isotropic diffuse emissions (fixing the spectral index of the Galactic component to 0). The normalization of 4FGL J0414.6$+$5711, located in the northwest of \src (at $\sim$ 2.4\dg from the SNR center), was also left free during the fit, while the spectral parameters of the other sources are fixed to their best-fit values found between 300 MeV and 3 TeV since they lie farther than 3.8\dg from the SNR center. We calculated a flux upper limit in the energy bands where the TS of the source is lower than 1. Figure~\ref{fig:SEDs} (left) shows the SEDs of \src and \srcbis, with the statistical and systematic errors. The emission at low energy is dominated by the contribution from \srcbis, while the emission from the SNR arises at higher energy. Figure~\ref{fig:SEDs} (right) compares the SEDs obtained with the Gaussian and the disk model, showing that the spatial model used for \src has a negligible impact on the spectral analysis. Since SNRs normally have a sharp edge (and the disk model is found to be spatially coincident with the radio morphology), we use the morphological and spectral properties of the disk throughout the paper to calculate the physical parameters of the SNR and discuss the nature of the emission.

%______________________________________________________________________________
\section{Multiwavelength observations}\label{sec:mwl}
%______________________________________________________________________________
\subsection{\hi spectrum and CO data towards \src}\label{sec:mwl_HI}

\begin{figure}[t!]
    \centering
    \includegraphics[scale=0.19]{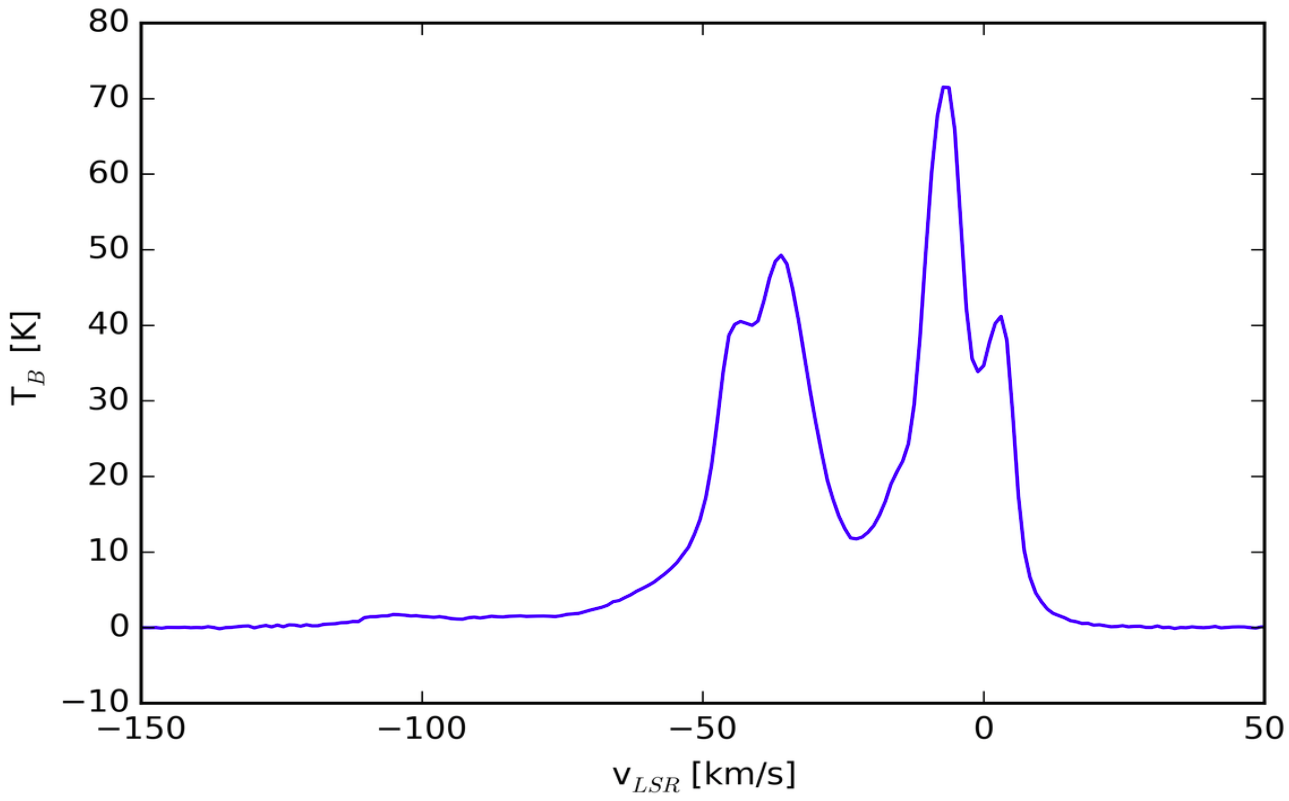}
    \caption{\hi spectrum obtained from the Leiden/Argentine/Bonn survey \citep{HI:2005} in the direction of \src.}
    \label{fig:HI_spec}
\end{figure}

Knowing the distance of \src is crucial in determining the physical size of the SNR and in turn understanding the origin of the $\gamma$-ray emission. Using data from the Leiden/Argentine/Bonn survey of Galactic \hi \citep{HI:2005}, we obtained the \hi spectrum in the direction of \src (shown in Figure~\ref{fig:HI_spec}) that displays clear velocity peaks at $-44.7$, $-35.9$, $-6.9$, and +2.9 km s$^{-1}$. The widths of all peaks are 5 km s$^{-1}$ or less, thus there is no evidence of shock-broadening from the SNR shock, where broadening is an indicator of the shock of an SNR overtaking nearby molecular clouds \citep{Wootten:1981}. Assuming a flat Galactic rotation curve with Galactocentric distance, $R_0$ = 8.34 kpc, and circular rotation speed at the sun $\Theta$ = 240 km s$^{-1}$, the three negative velocities noted above (a positive velocity is not permitted at this Galactic longitude) translate into distances of 4.83, 3.45, and 0.31 kpc, respectively. Since there is no hint of any interaction of the SNR with interstellar material, Figure~\ref{fig:HI_spec} is only indicative of the \hi emission towards \src and could be used to derive column density values in future works.

Looking at CO data \citep{Dame:2001}, we found molecular contents spatially coincident with the SNR (or at least part of it) with velocities between $-$9.8 km s$^{-1}$ and $-$3.8 km s$^{-1}$, which translate into distances of $d = 0.55_{-0.55}^{+0.65}$ kpc and $d = 0.01_{-0.01}^{+0.59}$ kpc, respectively.

\subsection{X-ray observations and distance estimate}\label{sec:mwl_rosat}

We used archival observations from the ROSAT all-sky survey \citep{Voges:1999} to determine whether there is significant X-ray emission (0.1$-$2.4 keV) in the direction of \src. The ROSAT data was extracted within a 1.3\dg circle centered on RA=4:27:08.528, Dec=+55:27:29.53 with a total exposure of 19 ks (observation ID: WG930708P\_N1\_SI01.N1, ROSAT/PSPC in survey mode). We used the tool \courrier{xselect} to extract the spectrum from the region of interest and the instrument response function (ancillary response function, ARF) was obtained using \gui{pcarf 2.1.3}, the latest version of the HEASOFT-FTOOL\footnote{\url{http://heasarc.gsfc.nasa.gov/ftools}} designed for ROSAT spectrum extraction. The background was derived using an off-source region, excluding any regions that seemed to contain point sources. No significant thermal or nonthermal emission is detected from \src. We modeled the emission using \courrier{XSPEC}\footnote{\url{https://heasarc.gsfc.nasa.gov/xanadu/xspec/}} \citep{XSPEC:1996} as an absorbed thermal non-equilibrium ionization (NEI) plasma \citep{Wilms:2000}. The Coulomb equilibration timescale is calculated following \cite{Vink:2011} and the initial electron temperature is assumed to be the higher of $T_p \times m_e/m_p$ or 0.3 keV \citep{Rakowski:2008}. We found a maximum absorbed photon flux of 0.0004 ph cm$^{-2}$ s$^{-1}$ and a corresponding maximum absorbed energy flux of $4.2 \times 10^{-13}$ erg cm$^{-2}$ s$^{-1}$ between 0.1 and 3.0 keV (with $kT_e = 0.2$ keV and $n_{e}t = 10^{11}$ s cm$^{-3}$). Using a power law with a spectral index $\Gamma$ = 2, the maximum  absorbed photon and energy fluxes are 0.0005 ph cm$^{-2}$ s$^{-1}$ and $1.1 \times 10^{-12}$ erg cm$^{-2}$ s$^{-1}$ between 0.1 and 3.0 keV. In the direction of \src, the maximum absorbing column density\footnote{Calculated with the Chandra Proposal Planning Toolkit \url{https://cxc.harvard.edu/toolkit/colden.jsp}} is $N_H = 3.97 \times 10^{21}$ cm$^{-2}$, giving maximum unabsorbed thermal and nonthermal photons$-$energy fluxes of 0.0171 ph cm$^{-2}$ s$^{-1}-8.6 \times 10^{-12}$ erg cm$^{-2}$ s$^{-1}$ ($kT_e = 0.2$ keV, $n_{e}t = 10^{11}$ s cm$^{-3}$) and 0.0063 ph cm$^{-2}$ s$^{-1}-3.6 \times 10^{-12}$ erg cm$^{-2}$ s$^{-1}$ ($\Gamma$ = 2).

To estimate the minimum distance to \src, we used the Sedov-Taylor self-similar solution
\begin{equation}\label{eq:Sedov_generic}
    R_s = \zeta \times \bigg ( \frac{E}{\rho_0} \bigg )^{1/5} t^{2/5} \hspace{0.5cm} \text{with} \hspace{0.5cm} \zeta = 1.152
,\end{equation}
where $R_s$, $E$, $\rho_0$, and $t$ are the radius of the SNR, the kinetic energy released by the supernova, the mass ambient density, and the age of the SNR, respectively. Equation~\ref{eq:Sedov_generic} can be written as
\begin{equation}\label{eq:Sedov_redim}
    R_s \approx 0.314 \times \bigg ( \frac{E_{\rm{51}}}{n_0} \bigg ) ^{1/5} t_{\rm{yr}}^{2/5} \hspace{0.2cm} \rm{pc}
,\end{equation}
where $E_{\rm{51}}$ is the kinetic energy released by the supernova, $n_0$ the ambient density, and $t_{\rm{yr}}$ the age of the SNR in units of 10$^{51}$ erg, cm$^{-3}$, and years. Assuming $E_{\rm{51}}$ = 1 and knowing the angular size of \src, we used different combinations of distance and age to calculate the corresponding ambient density. We then modeled the emission as an absorbed thermal NEI plasma for each combination of age, ambient, and column densities, and electron and proton temperature \citep{Castro:2011}. For each distance we thus calculated the corresponding column density, and we obtained an upper limit on the ambient density. The maximum ambient density allowed by ROSAT data is $n_0 = 3.6 \times 10^{-3}$ cm$^{-3}$. We also considered that an SNR at a declination of $\sim 55$\dg and younger than 1 kyr should have been reported in historical records. Thus, we imposed a lower limit on the age of the SNR of $t$ = 1 kyr, giving a minimum distance of $d$ = 0.7 kpc for an ambient density consistent with ROSAT data ($n_0 = 1.5 \times 10^{-3}$ cm$^{-3}$ at that distance). We note that the upper limit on the ambient density is not formally correct since an SNR with $E_{\rm{51}}$ = 1, $M_{\rm{ej}} \geq 1.4$ (the ejecta mass in units of solar mass) and $n_0 = 1.5 \times 10^{-3}$ cm$^{-3}$ has not yet entered the Sedov phase. This is, however, not critical for deriving an approximate upper limit on the ambient density. For a type Ia SNR, the Sedov phase is reached for $d$ = 0.9$-$1 kpc and $t$ = 3 kyr ($n_0 \sim 2.3 \times 10^{-3}$ cm$^{-3}$).

To constrain the maximum distance, we imposed a Mach number $\mathcal{M}^2 > 10$ (equivalent to a particle spectral index $p < 2.4$). With the sound speed $c_{s}^{2} = \frac{\gamma P_{\rm{ISM}}}{\rho_0}$ ($\gamma$ = 5/3 being the adiabatic index and $P_{\rm{ISM}}$ the interstellar medium pressure), this condition can be written as 
\begin{equation}
\rho_{0} v_{s}^{2} > 10 \rho_{0} c_{s}^{2} = 10 \gamma P_{\rm{ISM}}
.\end{equation}
In the Sedov phase, the shock velocity $v_s$ is 
\begin{equation}
v_{s} =  \frac{2}{5} \times \frac{R_s}{t}
,\end{equation}
with $R_s$ and $t$ the shock radius and the age of the SNR, respectively. Using Equation~\ref{eq:Sedov_generic}, this gives
\begin{equation}
\rho_{0} v_{s}^{2} = \bigg (\frac{2}{5} \bigg)^2 \zeta^5 \times \frac{E}{R_{s}^3}
.\end{equation}
The condition $\rho_{0} v_{s}^{2} > 10 \gamma P_{\rm{ISM}}$ translates into a relation among the physical radius of the SNR, the interstellar medium pressure, and the kinetic energy released by the supernova
\begin{equation}
    R_s < 1.69 \times \bigg ( \frac{P_{\rm{ISM}}}{k} \times \frac{1}{E} \bigg )^{-1/3} \hspace{0.2cm} \rm{kpc}
.\end{equation}

Assuming $E_{51}$ = 1 and taking $(P_{\rm{ISM}}/k) = 2.3n_0 T_{\rm{ISM}} = 3000$ K cm$^{-3}$, we found $R_s < 117.2$ pc, giving $d < 4.5$ kpc for $r$ = 1.497\dg. At this distance the age of the SNR is between $t$ = 161.5 kyr (using the maximum ambient density $n_0 = 3.6 \times 10^{-3}$ cm$^{-3}$ allowed by ROSAT data) and $t$ = 85.1 kyr (imposing a lower limit on the ambient density of $n_0 = 1 \times 10^{-3}$ cm$^{-3}$).

We note that the assumption $E_{51} = 1$ does not significantly impact the constraint on the ambient density since for $E_{51} = 0.1$ the maximum ambient density allowed by ROSAT data is $n_0 = 3.3 \times 10^{-3}$ cm$^{-3}$. The corresponding minimum and maximum distance estimates would be $d$ = 0.3 kpc and $d$ = 2.1 kpc.

In the following sections we assume $E_{51}$ = 1, and we use the minimum and maximum distance estimate $d$ = 0.7 kpc ($t$ = 1 kyr, $R_s$ = 18.3 pc, $n_0 = 1.5 \times 10^{-3}$ cm$^{-3}$) and $d$ = 4.5 kpc ($t$ = 85.1$-$161.5 kyr, $R_s$ = 117.2 pc, $n_0$ = (1.0$-$3.6) $\times 10^{-3}$ cm$^{-3}$) to discuss the origin of the emission.

%____________________________________________________________________________________________
\section{Discussion}\label{sec:discussion}
%____________________________________________________________________________________________
\subsection{Supernova remnant or pulsar wind nebula?}

The \gam data analysis presented here has led to the characterization of an extended $\gamma$-ray source whose centroid and extent are spatially coincident with the radio SNR. The broad size of the extended source and the correlation with the radio shell leaves few plausible scenarios for the nature of the \gam emission. Namely, the GeV emission can either arise from a pulsar wind nebula (PWN) or from an SNR.

The Gaussian morphology of the \gam emission and the hard \gam spectral index extending to hundreds of GeV make plausible the PWN scenario.
However, the \gam morphology does not get smaller at higher energies (Figure~\ref{fig:Edpt_morpho}) as expected from electron cooling, and ROSAT X-ray observations detect no significant nonthermal emission suggestive of a PWN in the direction of \src (Section~\ref{sec:mwl_rosat}). We nevertheless note that if the source is located at a large distance, the absorption of low-energy X-ray photons can prevent such a detection. The radio spectral index of the eastern shell \citep[$\alpha = -0.62 \pm 0.07$ and $\alpha=-0.40 \pm 0.17$,][]{Gerbrandt:2014, Gao:2014} and of the western shell \citep[$\alpha = -0.69 \pm 0.24$,][]{Gao:2014}, both spatially coincident with the $\gamma$-ray extent, are more compatible with that obtained for SNRs than for PWNe. Typical PWN radio spectral indices range from about $-0.3 \lesssim \alpha \lesssim 0$, while SNRs usually have a steeper radio spectral index $\alpha \lesssim -0.4$ \citep{Green:2017}. Thus, the measured radio spectral index disfavors a PWN scenario as the origin of the emission. Furthermore, there is no pulsar in the vicinity of the \gam emission that could power a PWN. The source \srcbis, located in the southern part of the SNR, has a pulsar-like spectrum and no pulsations were found with the  100-m Effelsberg radio telescope \citep{Barr:2013}. However, this potential pulsar could be radio quiet, or intensity variations from interstellar scintillation could have prevented detection at the epoch of observation. Therefore \srcbis may be a pulsar or another background source. Assuming that \srcbis was the compact remnant of the progenitor star that birthed \src, with an angular distance from the radio SNR center of 0.881\dg, it would have to be traveling with a velocity of $v_{\rm{PSR}}$ $\sim$ 10533 km s$^{-1}$ (taking $d$ = 0.7 kpc and $t$ = 1 kyr) or $v_{\rm{PSR}}$ $\sim$ 419$-$796 km s$^{-1}$ (taking $d$ = 4.5 kpc and $t$ = 161.5$-$85.1 kyr). Typical pulsar velocities range from $v_{\rm{PSR}}$ $\sim$ 400$-$500 km s$^{-1}$ \citep{Lyne:1995}, with extreme velocities exceeding 1000 km s$^{-1}$ \citep{Chatterjee:2005}. Thus, the association between \src and \srcbis would  only be possible when considering relatively far distances, such as $d$ = 4.5 kpc. From the spectral properties of \srcbis reported in Table~\ref{tab:spec}, we calculated an energy flux of $F_{0.1-100\hspace{0.1cm}\rm{GeV}} = 2.31$ $\times$ $10^{-11}$ erg cm$^{-2}$ s$^{-1}$, giving a luminosity $L_{0.1-100\hspace{0.1cm}\rm{GeV}} = 5.60$ $\times$ $10^{34} (d/4.5\hspace{0.1cm} \rm{kpc})^2$ erg s$^{-1}$. The empirical relation between the luminosity and the spin-down power of the pulsar $L_{0.1-100\hspace{0.1cm}\rm{GeV}} = \sqrt{10^{33} \dot{E}}$ erg s$^{-1}$ \citep{Abdo:2013} gives $\dot{E}$ = 3.14 $\times$ $10^{36}$ $(d/4.5\hspace{0.1cm} \rm{kpc})^4$ erg s$^{-1}$ (taking a beam correction factor of 1), indicating that \srcbis could be an energetic pulsar. However, if \srcbis is associated with the SNR and powers a PWN responsible for the observed GeV emission, the PWN $\gamma$-ray extent would be as large as the radio extent of its host SNR which never occurs in composite systems. All these arguments disfavor a PWN scenario as the origin of the main \gam emission, keeping in mind that a PWN located in the line of sight and contributing to a part of the $\gamma$-ray emission cannot be ruled out.

The spatial correlation between the radio and the \gam emissions, together with the lack of a \gam morphology shrinking at higher energies, points toward an SNR scenario. This is supported by the radio shell-like appearance, the nonthermal radio spectrum similar to that obtained in SNRs, and the detection of red optical filaments structures by \cite{Gerbrandt:2014}. Unless a pulsar is detected in the future that could power a PWN contributing to the GeV emission, we argue that the \gam emission is likely produced by the SNR \src and is the counterpart of the radio emission detected by \cite{Gao:2014}.

\subsection{Evolutionary stage of \src}
    
\begin{figure}[th!]
    \centering
    \includegraphics[scale=0.33]{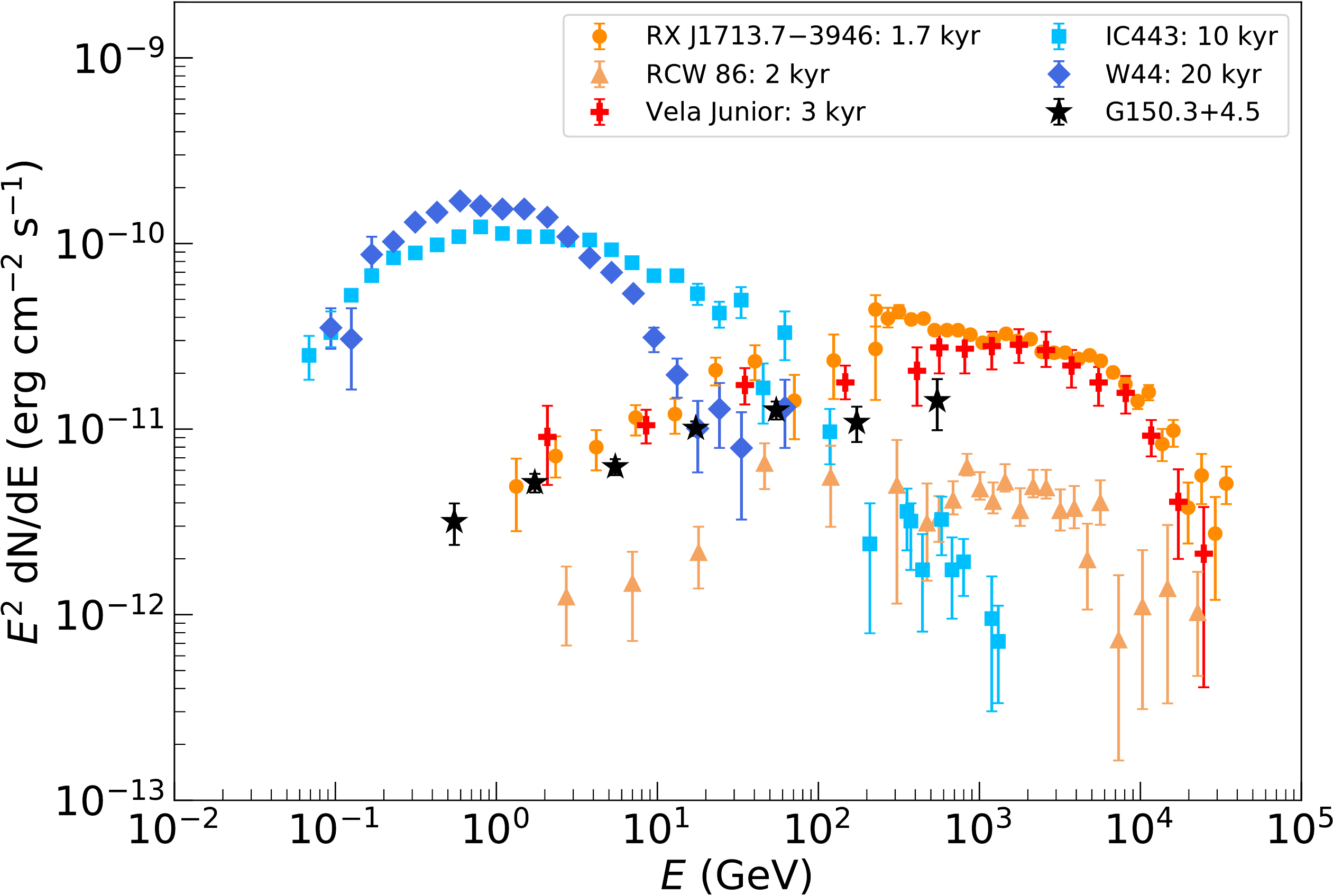}
    \caption{SEDs of several \fermi SNRs interacting with molecular clouds (blue tone colors) and being dynamically young (orange tone colors). The black stars correspond to the SED of \src shown in Figure~\ref{fig:SEDs} (right) using the disk model. References: RX J1713.7$-$3946 \citep{RXJ:2018}, RCW 86 \citep{RCW86:2016}, Vela Junior \citep{VelaJr:2018}, IC 443, and W44 \citep{Ackermann:2013}.}
    \label{fig:SEDs_SNRs}
\end{figure}

In the following sections we assume that the entire \gam emission comes from the SNR \src. We place the SNR in context within the current population of \fermi SNRs to assess its evolutionary stage. Figure~\ref{fig:SEDs_SNRs} shows the \fermi SED of \src (obtained in Section~\ref{sec:fermi_spec} using a uniform disk) overlaid on the spectra of several \fermi observed SNRs with ages ranging from $\sim$ 1 to 20 kyr. As shown in Figure~\ref{fig:SEDs_SNRs}, spectral breaks are commonly observed below a few GeV in SNRs interacting with nearby molecular material. \src exhibits a hard spectrum extending to hundreds of GeV with no spectral break, and is thus spectrally similar to the dynamically young and shell-type SNRs like RX J1713.7$-$3946 or Vela Junior. 

\begin{figure}[th!]
    \centering
    \includegraphics[scale=0.33]{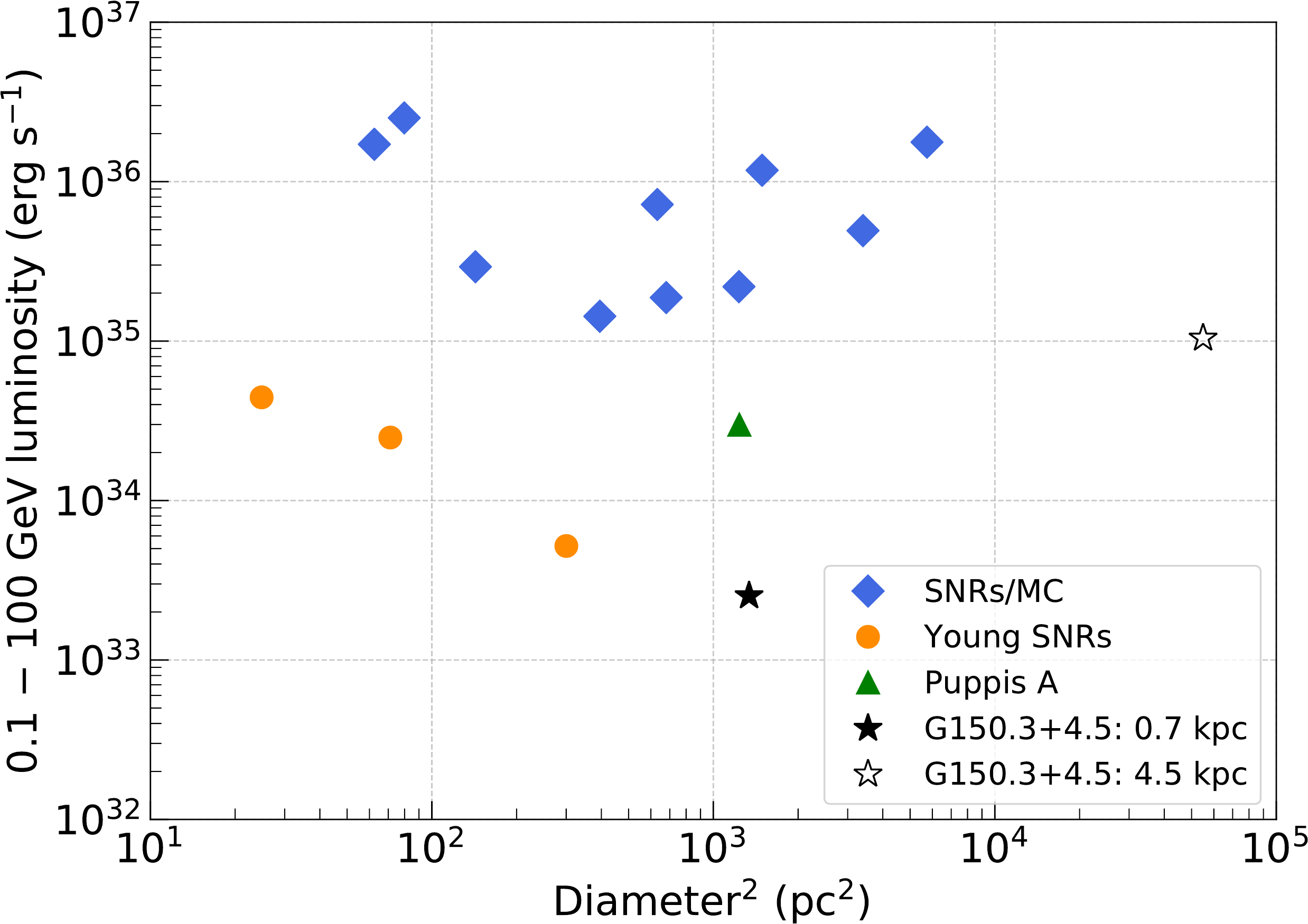}
    \caption{Luminosity (0.1$-$100 GeV) of several \fermi observed SNRs plotted against their diameter squared. Blue diamonds are the SNRs interacting with molecular clouds, while orange circles are young SNRs. The SNR Puppis A, a middle-aged SNR that is not interacting with a molecular cloud, is represented by the green triangle. Black stars correspond to the SNR \src using a distance of 0.7 kpc and 4.5 kpc, respectively. The figure is reproduced from \cite{Thompson:2012} with the addition of \src.}
    \label{fig:GeV_lum}
\end{figure}

Figure~\ref{fig:GeV_lum} depicts the luminosity of several \fermi SNRs with respect to their diameter squared. Between 100 MeV to 100 GeV, \src has a luminosity of $2.53 \times 10^{33} (d/0.7\hspace{0.1cm} \rm{kpc})^2$ erg s$^{-1}$. For the minimum distance, the luminosity of \src is low and similar to that obtained for young SNRs, while for the maximum distance the luminosity is  closer to those of SNRs interacting with molecular clouds (MCs). However, there is no hint of an interaction between \src and a MC in the radio data explored by \cite{Gerbrandt:2014} and \cite{Gao:2014}, as in the catalog of molecular clouds \citep{Rice:2016} and of Galactic \hii regions \citep{Wise:2014}, which do not report any object close to \src. The hard spectral shape of \src and its likely low luminosity from 100 MeV to 100 GeV supports the dynamically young and non-interacting SNR scenario, and therefore a near distance.

\subsection{Broadband nonthermal modeling}\label{sec:modeling}

Particle acceleration arises at the shock front of the SNR \src, producing $\gamma$ rays up to hundreds of GeV. These $\gamma$ rays can  either be produced by accelerated electrons through IC scattering on photon fields and Bremsstrahlung, or by accelerated protons colliding with ambient matter and leading to the decay of the neutral pion. To understand the origin of the emission, we performed multiwavelength modeling using the \courrier{naima} package \citep{Zabalza:2015} in a one-zone model assumption. As shown in Figure~\ref{fig:SEDs} (right), the choice of the spatial model for \src has no impact on the spectral analysis. We used the radio fluxes derived in \cite{Gerbrandt:2014} and the $\gamma$-ray spectrum obtained with the disk model (Section~\ref{sec:fermi_morpho}), and we factored the X-ray absorption into the model, allowing the use of the robust upper limit on the absorbed X-ray flux obtained with ROSAT data (Section~\ref{sec:mwl_rosat}). Since the radio fluxes were derived in a region of $\sim$ 2.14\dg $\times$ 0.63\dg \citep{Gerbrandt:2014} that does not encompass the entire radio SNR \cite[$\sim$ 3\dg $\times$ 2\dg, as reported in][]{Gao:2014}, we considered these data points as lower limits. We modeled the broadband nonthermal emission of the SNR assuming $E_{51}$ = 1 and for the two extreme distances $d$ = 0.7 kpc and $d$ = 4.5 kpc.

We described the proton spectrum by a power law with an exponential  cutoff (defining the maximum energy reached by the particles) and the electron spectrum by a broken power law with an exponential cutoff. The break in the electron spectrum is due to synchrotron cooling, and occurs at the energy $E_b$ above which the particle spectral index is $s_{\rm{e,2}}$ = $s_{\rm{e,1}} + 1$, with $s_{\rm{e,1}}$ being the spectral index below the energy break. To be consistent with the measured radio spectral index ($\alpha = -0.38$ $\pm$ 0.10), we fixed the electron spectral index to $s_{\rm{e,1}}$ = 1.8, giving $s_{\rm{e,2}}$ = 2.8, and we used the same value for the proton spectral index $s_{\rm{p}} = 1.8$. To constrain the energy break and the particle maximum energy, we considered the synchrotron loss time 
\begin{equation}
    \tau_{\rm{sync}} = (1.25 \times 10^{3}) \times E_{\rm{TeV}}^{-1} B_{\rm{100}}^{-2} \hspace{0.2cm} \rm{yr}
\end{equation}
and the acceleration timescale
\begin{equation}\label{eq:emax}
    t_{\rm{acc}} = 30.6 \times \frac{3r^2}{16(r-1)} \times k_{0}(E) \times E_{\rm{TeV}} B_{\rm{100}}^{-1} v_{\rm{s, 3}}^{-2} \hspace{0.2cm} \rm{yr} 
,\end{equation}
where $r$ is the shock compression ratio, $k_{0}$ is the ratio between the mean free path and the gyroradius (equivalent to $B_{\rm{tot}}^2/B_{\rm{turb}}^2$), and $B_{\rm{100}}$ and $v_{\rm{s, 3}}$ are the magnetic field and the shock velocity in units of 100 $\mu$G and $1000$ km s$^{-1}$, respectively \citep{Parizot:2006}. For young SNRs we expect $k_{0}$ $\sim$ 1, while for evolved systems we expect $k_{0} >$ 1. The break and maximum energy of the electrons are calculated equating $\tau_{\rm{sync}}$ = $t_{\rm{age}}$ and $t_{\rm{acc}} = \rm{min}(t_{\rm{age}}, \tau_{\rm{sync}}$), respectively, while the maximum energy of the protons is found with $t_{\rm{age}} = t_{\rm{acc}}$. For the IC scattering on photon fields, we considered the cosmic microwave background (U$_{\rm{CMB}}$ = 0.26 eV cm$^{-3}$, $T_{\rm{CMB}}$ = 2.7 K), the infrared and optical emissions for which the temperature and energy density are estimated with the GALPROP code\footnote{\url{https://galprop.stanford.edu}} \citep{Porter:2008}. The total energy of accelerated particles is integrated above 1 GeV.

\begin{table*}[t!]
\footnotesize{
    \centering
    \begin{tabular}{l|cccccccccc}
    \hline
    \hline
$d$ (kpc) / $t$ (kyr)      & $B$ ($\mu$G)  & $W_{\rm{p}}$ (erg) & $K_{\rm{ep}}$ & $s_{\rm{e,1}} = s_{\rm{e,2}}$ & $s_{\rm{p}}$ & $n_0$ (cm$^{-3}$) &  $E_{\rm{max, e}}$ (TeV) & $E_{\rm{max, p}}$ (TeV) & $k_0$ & $v_s$ (km s$^{-1}$) \\
\hline
0.7 / 1.0  & 5 & 10$^{50}$ &  1 $\times$ 10$^{-3}$ & 1.8* & 1.8 & 1.5 $\times$ $10^{-3}$ &  5.2 & 5.2 & 16 & 7163\\
4.5 / 85.1   &  5 & 10$^{50}$ & 5 $\times$ 10$^{-2}$ & 1.8* & 1.8 & 1.0 $\times$ $10^{-3}$  &  5.8  & 5.8 & 7 & 539 \\
\hline
    \end{tabular}
    \caption{Values used for the broadband nonthermal modeling shown in Figure~\ref{fig:SED_model}. Asterisks denote the values constrained by radio data. The particle maximum energy values  ($E_{\rm{max, e}}$ and $E_{\rm{max, p}}$) are calculated using the values of $B$, $k_0$, and the shock velocity $v_s$ derived in the Sedov model.}
    \label{tab:SED_model}
}
\end{table*}
\begin{figure*}[t!]
    \includegraphics[scale=0.33]{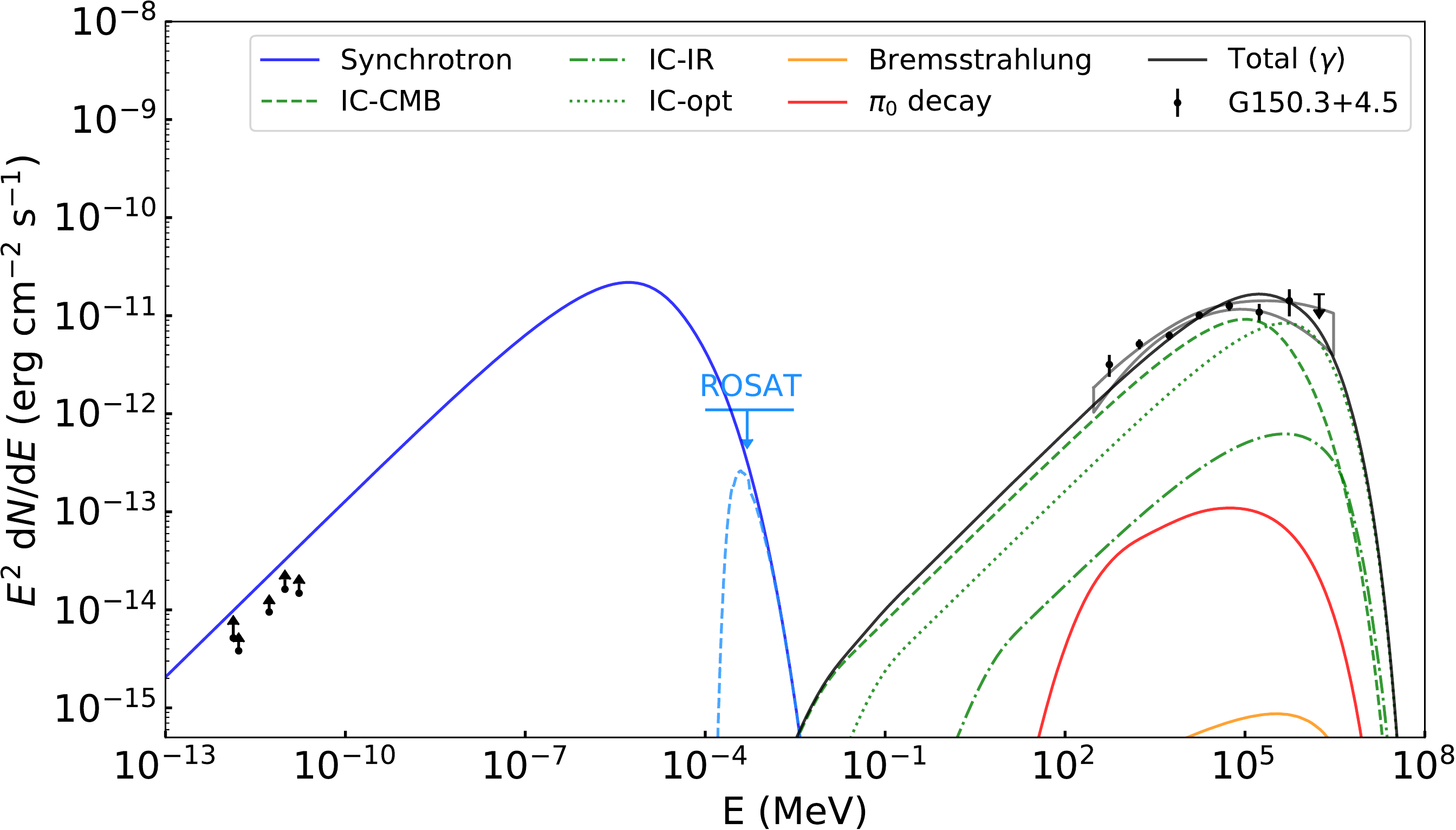}
    \includegraphics[scale=0.33]{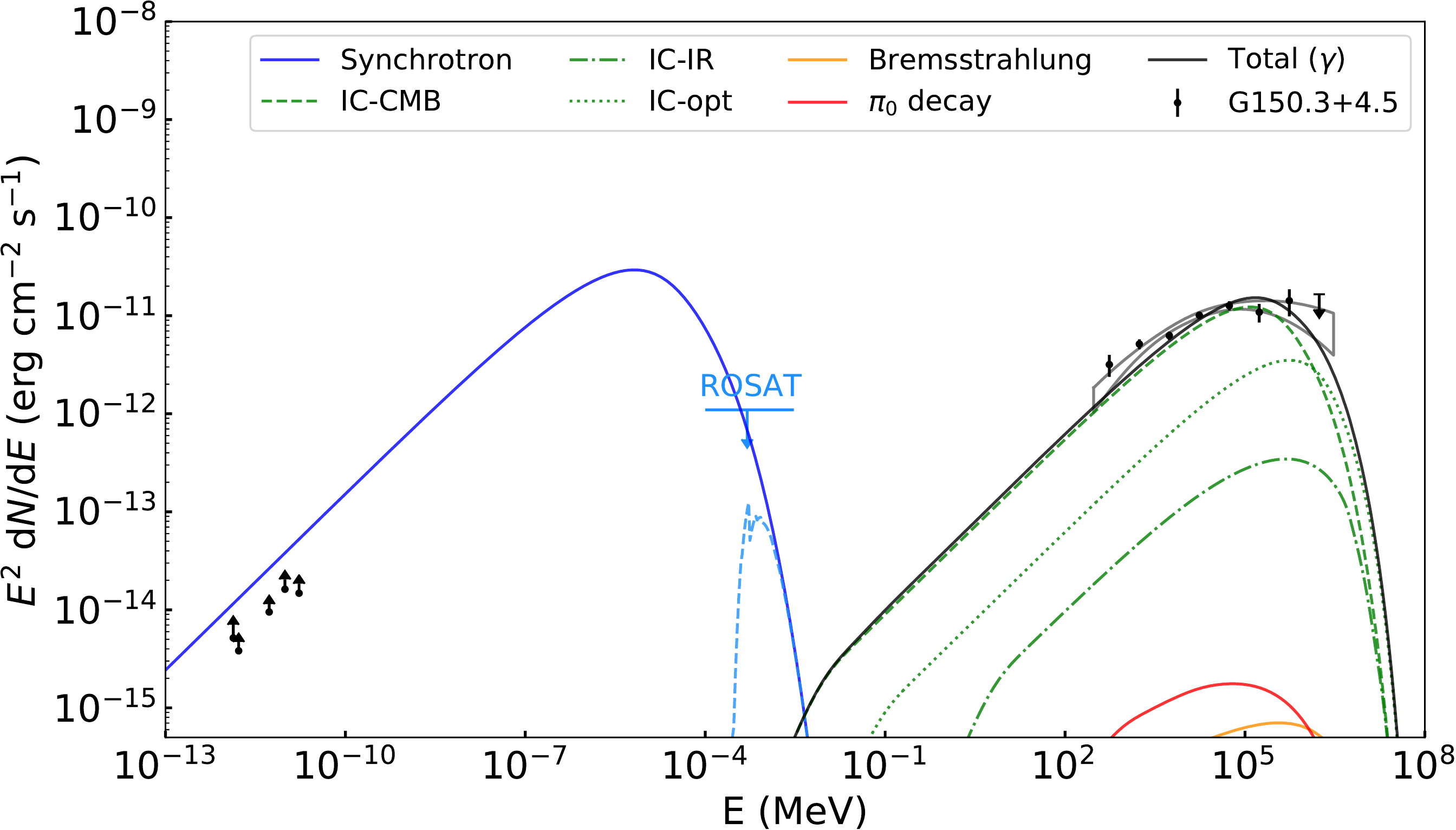}
    \caption{Broadband nonthermal modeling of the SNR \src in a leptonic scenario using a distance of $d$ = 0.7 kpc (left) and $d$ = 4.5 kpc (right). The radio fluxes, which are near lower limits (see Section~\ref{sec:modeling}), come from \cite{Gerbrandt:2014} and the best-fit spectrum and $\gamma$-ray fluxes are those derived using the disk model. The ROSAT upper limit on the absorbed flux is derived in Section~\ref{sec:mwl_rosat} and has to be compared to the absorbed X-ray synchrotron model (dashed blue curve). The values used for the plots are reported in Table~\ref{tab:SED_model}.}
    \label{fig:SED_model}
\end{figure*}

For a distance of $d$ = 0.7 kpc, the temperature and energy density of the infrared and optical emissions are $T_{\rm{IR}}$ = 25.2 K, U$_{\rm{IR}}$ = 0.34 eV cm$^{-3}$ and $T_{\rm{opt}}$ = 2004.9 K, U$_{\rm{opt}}$ = 0.52 eV cm$^{-3}$. We used the ambient density of $n_0 = 1.5 \times 10^{-3}$ cm$^{-3}$ derived in Section~\ref{sec:mwl_rosat}, and the column density of $N_H$ = 5.23 $\times$ $10^{20}$ cm$^{-2}$ (corresponding to a distance of $d$ = 0.7 kpc) to model the absorbed X-ray nonthermal emission. The radio fluxes constrain the magnetic field to be relatively low. With $W_{\rm{p}} = 10^{50}$ erg (10\% of $E_{51}$ going into CR protons), the data can be reproduced with a downstream magnetic field of $B = 5$ $\mu$G and an electron-to-proton ratio of $K_{\rm{ep}} = 1 \times 10^{-3}$. We used the value of the velocity in the Sedov phase $v_{\rm{s}} = 7163$ km s$^{-1}$, which is consistent with the data for $E_{\rm{b}}$ = $E_{\rm{max, e}}$ = $E_{\rm{max, p}}$ = 5.2 TeV implying $k_0$ = 16 (using $r = 4$). Given the low magnetic field and age of the SNR, electrons do not suffer significantly from synchrotron cooling so there is no break in the electron spectrum. We note that if the SNR is in the ejecta-dominated phase, the velocity should be even higher than $v_{\rm{s}} = 7163$ km s$^{-1}$. Although the value of $k_0$ is unknown, $k_0$ = 16 is higher than one would expect. Since we have $E_{\rm{max}} \propto v_s^{2}/k_0$, a lower value of $k_0$ would imply a reduced shock velocity to still fit the data. That could indicate that the SNR is located at a larger distance than 0.7 kpc (since we have $v_{\rm{s}} \propto R_{s}^{-3}$). Table~\ref{tab:SED_model} reports the physical parameters used for the broadband modeling of the SNR spectrum, shown in Figure~\ref{fig:SED_model} (left). 

For a distance of $d$ = 4.5 kpc, the temperature and the energy density of the infrared and optical emissions are $T_{\rm{IR}}$ = 25.2 K, U$_{\rm{IR}}$ = 0.11 eV cm$^{-3}$ and $T_{\rm{opt}}$ = 2004.9 K, U$_{\rm{opt}}$ = 0.24 eV cm$^{-3}$. We first discuss the case of $t$ = 85.1 kyr obtained with $n_0 = 1.0 \times 10^{-3}$ cm$^{-3}$ and we used a column density of $N_H$ = 2.63 $\times$ $10^{21}$ cm$^{-2}$ for the absorbed X-ray synchrotron model (corresponding to $d$ = 4.5 kpc). Setting $W_{\rm{p}} = 10^{50}$ erg and using a shock velocity of $v_{s}$ = 539 km s$^{-1}$ calculated in the Sedov model, the data can be reproduced with $B = 5$ $\mu$G, $K_{\rm{ep}} = 5 \times 10^{-2}$, and $E_{\rm{b}}$ = $E_{\rm{max, e}}$ = $E_{\rm{max, p}}$ = 5.8 TeV obtained using $k_0$ = 7. The spectrum is shown in Figure~\ref{fig:SED_model} (right) with the corresponding values reported in Table~\ref{tab:SED_model}. For $t$ = 161.5 kyr, the shock velocity in the Sedov phase is $v_{s}$ = 284 km s$^{-1}$ implying a lower value of $k_0$ to still fit the data. We also note that the value of $K_{\rm{ep}} = 5 \times 10^{-2}$ is slightly higher than one would expect, strengthening the fact that the distance of \src is likely smaller than $d$ = 4.5 kpc.

Using the two extreme distances of $d$ = 0.7 kpc and $d$ = 4.5 kpc, the broadband spectrum of the SNR is explained by a leptonic scenario, with acceleration of particles up to $\sim$ 5 TeV and a downstream magnetic field of $B$ = 5 $\mu$G. Under standard shock conditions, this gives an interstellar magnetic field of $B_{\rm{ISM}} = 1.5$ $\mu$G (taking the isotropic compression ratio with $r$ = 4). Since the radio fluxes derived in \cite{Gerbrandt:2014} are likely underestimated, we used them as near lower limits. In each of our models, the synchrotron flux never exceeds $\sim$ 2$-$3 times the values of the radio data, which is reasonable given the difference between the extraction region size taken in \cite{Gerbrandt:2014} and the extent of the radio SNR (assuming a uniform emission). The minimum and maximum distances  respectively require a higher value of $k_0$ and $K_{\rm{ep}}$ than one would expect, which indicate that the SNR is indeed located between 0.7 kpc and 4.5 kpc. We did not consider the hadron-dominated scenario due to the low maximum ambient density allowed by ROSAT data (Section~\ref{sec:mwl_rosat}), which would lead to an unrealistic value of $W_{\rm{p}}$, and due to the lack of any hint of interaction of the SNR with interstellar material (Section~\ref{sec:mwl_HI}). For the same reason, we also note that the SNR is likely not located at the same distance as those of the enhanced \hi and CO emissions (derived in Section~\ref{sec:mwl_HI}).

Deeper multiwavelength observations are necessary to better constrain the broadband emission from \src. That concerns radio observations to derive the synchrotron spectrum in a region encompassing the entire SNR, X-ray observations to obtain stronger upper limits on the thermal emission and very high-energy observations with Cherenkov telescopes to assess the maximum energy reached by particles in this Galactic cosmic-ray accelerator. Finally, a distance estimate would allow us to more precisely determine the evolutionary stage of the SNR \src.

%____________________________________________________________________________________________
\section{Conclusions}
%____________________________________________________________________________________________
We analyzed more than 10 years of \fermi data and we investigated the morphological and the spectral properties of the $\gamma$-ray emission towards the recently detected radio SNR \src. From 1 GeV to 3 TeV the emission is adequately described by a 2D symmetric Gaussian or a disk model, which is spatially coincident with the radio emission. Going down to 300 MeV, the spectrum is best described by a logarithmic parabola with a hard spectral index $\alpha$ = 1.62 $\pm$ 0.04$_{\rm{stat}}$ $\pm$ 0.23$_{\rm{syst}}$ at $E_{0}$ = 9.0 GeV. The point source \srcbis, located in the southern part of the SNR, has a pulsar-like spectrum and dominates the low-energy part of the $\gamma$-ray emission ($E <$ 3 GeV), while the contribution from the SNR arises at higher energy. 

We did not detect significant thermal and nonthermal X-ray emission using ROSAT all-sky survey data, which implies a maximum ambient density $n_0 = 3.6 \times 10^{-3}$ cm$^{-3}$. Setting a lower limit on the age of the SNR $t$ = 1 kyr, we estimated a minimum distance of $d$ = 0.7 kpc, which is consistent with the ROSAT upper limit on the ambient density. Using physical considerations, we estimated a maximum distance of $d$ = 4.5 kpc. Using the maximum ambient density allowed by ROSAT data and a lower limit of $n_0 = 0.1 \times 10^{-3}$, the age of the SNR is $t$ = 85.1$-$161.5 kyr.

The association between \src and the pulsar-like \srcbis is unclear and would be possible only if the SNR is located at relatively far distances, giving \srcbis a reasonable transverse velocity. Since there is no hint for a PWN as the origin of the $\gamma$-ray emission, the latter being spatially coincident with the radio SNR, we argued that the GeV emission is likely produced by the SNR \src. We compared the spectrum and luminosity of \src with those of other \fermi observed SNRs, including young systems and evolved SNRs interacting with molecular clouds. We found that \src is spectrally similar to the dynamically young and shell-type \fermi SNRs. We modeled the broadband nonthermal emission of \src with a leptonic scenario that implies a downstream magnetic field of $B = 5$ $\mu$G and an acceleration of particles up to few TeV energies.

Deeper multiwavelength observations are required to better constrain the distance, the environment of \src and its synchrotron spectrum. Since the detection of X-ray emission with \ita{Chandra} or \ita{XMM-Newton} can be challenging for sources with an angular size of 3\dg, large field-of-view X-ray instruments like eROSITA would be helpful to assess any thermal and nonthermal emission towards \src. Very high-energy analyses are promising in constraining the highest part of the $\gamma$-ray spectrum. Although not detected with 507 days of observations by HAWC \citep{2HWC:2017}, \src is a good candidate to be observed with the next generation of Cherenkov telescopes CTA (Cherenkov Telescope Array) that will give clear insights into the maximum energy reached by particles in this dynamically young SNR. Finally, deeper pulsation searches on \srcbis would help investigate its nature and its possible association with \src.

\small{
\begin{acknowledgements}
We thank the referee for the helpful comments on this paper, and X. Y. Gao and J. L. Han for providing us the Urumqi 6 cm map of the SNR. JD, MLG and MHG acknowledge support from Agence Nationale de la Recherche (grant ANR- 17-CE31-0014). The Fermi LAT Collaboration acknowledges generous ongoing support from a number of agencies and institutes that have supported both the development and the operation of the LAT as well as scientific data analysis. These include the National Aeronautics and Space Administration and the Department of Energy in the United States, the Commissariat à l’Energie Atomique and the Centre National de la Recherche Scientifique / Institut National de Physique Nucléaire et de Physique des Particules in France, the Agenzia Spaziale Italiana and the Istituto Nazionale di Fisica Nucleare in Italy, the Ministry of Education, Culture, Sports, Science and Technology (MEXT), High Energy Accelerator Research Organization (KEK) and Japan Aerospace Exploration Agency (JAXA) in Japan, and the K. A. Wallenberg Foundation, the Swedish Research Council and the Swedish National Space Board in Sweden. Additional support for science analysis during the operations phase is gratefully acknowledged from the Istituto Nazionale di Astrofisica in Italy and the Centre National d’Études Spatiales in France. This work performed in part under DOE Contract DE-AC02-76SF00515.
\end{acknowledgements}}

\bibliographystyle{aa}
\bibliography{SNRG150}

\begin{thebibliography}{39}
\expandafter\ifx\csname natexlab\endcsname\relax\def\natexlab#1{#1}\fi

\bibitem[{{Abdo} {et~al.}(2013){Abdo}, {Ajello}, {Allafort}, {Baldini},
  {Ballet}, {Barbiellini}, {Baring}, {Bastieri}, {Belfiore}, {Bellazzini}, \&
  et~al.}]{Abdo:2013}
{Abdo}, A.~A., {Ajello}, M., {Allafort}, A., {et~al.} 2013, \apjs, 208, 17

\bibitem[{{Abeysekara} {et~al.}(2017){Abeysekara}, {Albert}, {Alfaro},
  {Alvarez}, {{\'A}lvarez}, {Arceo}, {Arteaga-Vel{\'a}zquez}, {Ayala Solares},
  {Barber}, {Baughman}, {Bautista-Elivar}, {Becerra Gonzalez}, {Becerril},
  {Belmont-Moreno}, {BenZvi}, {Berley}, {Bernal}, {Braun}, {Brisbois},
  {Caballero-Mora}, {Capistr{\'a}n}, {Carrami{\~n}ana}, {Casanova}, {Castillo},
  {Cotti}, {Cotzomi}, {Couti{\~n}o de Le{\'o}n}, {de la Fuente}, {De Le{\'o}n},
  {Diaz Hernandez}, {Dingus}, {DuVernois}, {D{\'{\i}}az-V{\'e}lez},
  {Ellsworth}, {Engel}, {Fiorino}, {Fraija}, {Garc{\'{\i}}a-Gonz{\'a}lez},
  {Garfias}, {Gerhardt}, {Gonz{\'a}lez Mu{\~n}oz}, {Gonz{\'a}lez}, {Goodman},
  {Hampel-Arias}, {Harding}, {Hernandez}, {Hernandez-Almada}, {Hinton}, {Hui},
  {H{\"u}ntemeyer}, {Iriarte}, {Jardin-Blicq}, {Joshi}, {Kaufmann}, {Kieda},
  {Lara}, {Lauer}, {Lee}, {Lennarz}, {Le{\'o}n Vargas}, {Linnemann},
  {Longinotti}, {Raya}, {Luna-Garc{\'{\i}}a}, {L{\'o}pez-Coto}, {Malone},
  {Marinelli}, {Martinez}, {Martinez-Castellanos}, {Mart{\'{\i}}nez-Castro},
  {Mart{\'{\i}}nez-Huerta}, {Matthews}, {Miranda-Romagnoli}, {Moreno},
  {Mostaf{\'a}}, {Nellen}, {Newbold}, {Nisa}, {Noriega-Papaqui}, {Pelayo},
  {Pretz}, {P{\'e}rez-P{\'e}rez}, {Ren}, {Rho}, {Rivi{\`e}re},
  {Rosa-Gonz{\'a}lez}, {Rosenberg}, {Ruiz-Velasco}, {Salazar}, {Salesa Greus},
  {Sandoval}, {Schneider}, {Schoorlemmer}, {Sinnis}, {Smith}, {Springer},
  {Surajbali}, {Taboada}, {Tibolla}, {Tollefson}, {Torres}, {Ukwatta},
  {Vianello}, {Villase{\~n}or}, {Weisgarber}, {Westerhoff}, {Wisher}, {Wood},
  {Yapici}, {Younk}, {Zepeda}, \& {Zhou}}]{2HWC:2017}
{Abeysekara}, A.~U., {Albert}, A., {Alfaro}, R., {et~al.} 2017, \apj, 843, 40

\bibitem[{{Abeysekara} {et~al.}(2020){Abeysekara}, {Archer}, {Benbow}, {Bird},
  {Brose}, {Buchovecky}, {Buckley}, {Chromey}, {Cui}, {Daniel}, {Das},
  {Dwarkadas}, {Falcone}, {Feng}, {Finley}, {Fortson}, {Gent}, {Gilland ers},
  {Giuri}, {Gueta}, {Hanna}, {Hassan}, {Hervet}, {Holder}, {Hughes},
  {Humensky}, {Kaaret}, {Kar}, {Kelley-Hoskins}, {Kertzman}, {Kieda}, {Krause},
  {Krennrich}, {Kumar}, {Lang}, {Maier}, {Moriarty}, {Mukherjee},
  {Nievas-Rosillo}, {O'Brien}, {Ong}, {Park}, {Petrashyk}, {Pfrang}, {Pohl},
  {Pueschel}, {Quinn}, {Ragan}, {Reynolds}, {Richards}, {Roache}, {Sadeh},
  {Santander}, {Sembroski}, {Shahinyan}, {Sushch}, {Weinstein}, {Wilcox},
  {Wilhelm}, {Williams}, {Williamson}, {Zitzer}, \&
  {Ghiotto}}]{Abeysekara_CasA:2020}
{Abeysekara}, A.~U., {Archer}, A., {Benbow}, W., {et~al.} 2020, \apj, 894, 51

\bibitem[{{Acero} {et~al.}(2016){Acero}, {Ackermann}, {Ajello}, {Baldini},
  {Ballet}, {Barbiellini}, {Bastieri}, {Bellazzini}, {Bissaldi}, {Blandford},
  {Bloom}, {Bonino}, {Bottacini}, {Brand t}, {Bregeon}, {Bruel}, {Buehler},
  {Buson}, {Caliandro}, {Cameron}, {Caputo}, {Caragiulo}, {Caraveo}, {Casand
  jian}, {Cavazzuti}, {Cecchi}, {Chekhtman}, {Chiang}, {Chiaro}, {Ciprini},
  {Claus}, {Cohen}, {Cohen-Tanugi}, {Cominsky}, {Condon}, {Conrad}, {Cutini},
  {D'Ammando}, {de Angelis}, {de Palma}, {Desiante}, {Digel}, {Di Venere},
  {Drell}, {Drlica-Wagner}, {Favuzzi}, {Ferrara}, {Franckowiak}, {Fukazawa},
  {Funk}, {Fusco}, {Gargano}, {Gasparrini}, {Giglietto}, {Giommi}, {Giordano},
  {Giroletti}, {Glanzman}, {Godfrey}, {Gomez-Vargas}, {Grenier}, {Grondin},
  {Guillemot}, {Guiriec}, {Gustafsson}, {Hadasch}, {Harding}, {Hayashida},
  {Hays}, {Hewitt}, {Hill}, {Horan}, {Hou}, {Iafrate}, {Jogler},
  {J{\'o}hannesson}, {Johnson}, {Kamae}, {Katagiri}, {Kataoka}, {Katsuta},
  {Kerr}, {Kn{\"o}dlseder}, {Kocevski}, {Kuss}, {Laffon}, {Lande}, {Larsson},
  {Latronico}, {Lemoine-Goumard}, {Li}, {Li}, {Longo}, {Loparco}, {Lovellette},
  {Lubrano}, {Magill}, {Maldera}, {Marelli}, {Mayer}, {Mazziotta}, {Michelson},
  {Mitthumsiri}, {Mizuno}, {Moiseev}, {Monzani}, {Moretti}, {Morselli},
  {Moskalenko}, {Murgia}, {Nemmen}, {Nuss}, {Ohsugi}, {Omodei}, {Orienti},
  {Orlando}, {Ormes}, {Paneque}, {Perkins}, {Pesce-Rollins}, {Petrosian},
  {Piron}, {Pivato}, {Porter}, {Rain{\`o}}, {Rando}, {Razzano}, {Razzaque},
  {Reimer}, {Reimer}, {Renaud}, {Reposeur}, {Rousseau}, {Saz Parkinson},
  {Schmid}, {Schulz}, {Sgr{\`o}}, {Siskind}, {Spada}, {Spandre}, {Spinelli},
  {Strong}, {Suson}, {Tajima}, {Takahashi}, {Tanaka}, {Thayer}, {Thompson},
  {Tibaldo}, {Tibolla}, {Torres}, {Tosti}, {Troja}, {Uchiyama}, {Vianello},
  {Wells}, {Wood}, {Wood}, {Yassine}, {den Hartog}, \& {Zimmer}}]{SNR:2016}
{Acero}, F., {Ackermann}, M., {Ajello}, M., {et~al.} 2016, \apjs, 224, 8

\bibitem[{{Acero} {et~al.}(2015){Acero}, {Lemoine-Goumard}, {Renaud}, {Ballet},
  {Hewitt}, {Rousseau}, \& {Tanaka}}]{Acero:2015}
{Acero}, F., {Lemoine-Goumard}, M., {Renaud}, M., {et~al.} 2015, \aap, 580, A74

\bibitem[{{Ackermann} {et~al.}(2013){Ackermann}, {Ajello}, {Allafort},
  {Baldini}, {Ballet}, {Barbiellini}, {Baring}, {Bastieri}, {Bechtol},
  {Bellazzini}, {Blandford}, {Bloom}, {Bonamente}, {Borgland}, {Bottacini},
  {Brandt}, {Bregeon}, {Brigida}, {Bruel}, {Buehler}, {Busetto}, {Buson},
  {Caliandro}, {Cameron}, {Caraveo}, {Casandjian}, {Cecchi}, {{\c C}elik},
  {Charles}, {Chaty}, {Chaves}, {Chekhtman}, {Cheung}, {Chiang}, {Chiaro},
  {Cillis}, {Ciprini}, {Claus}, {Cohen-Tanugi}, {Cominsky}, {Conrad}, {Corbel},
  {Cutini}, {D'Ammando}, {de Angelis}, {de Palma}, {Dermer}, {do Couto e
  Silva}, {Drell}, {Drlica-Wagner}, {Falletti}, {Favuzzi}, {Ferrara},
  {Franckowiak}, {Fukazawa}, {Funk}, {Fusco}, {Gargano}, {Germani},
  {Giglietto}, {Giommi}, {Giordano}, {Giroletti}, {Glanzman}, {Godfrey},
  {Grenier}, {Grondin}, {Grove}, {Guiriec}, {Hadasch}, {Hanabata}, {Harding},
  {Hayashida}, {Hayashi}, {Hays}, {Hewitt}, {Hill}, {Hughes}, {Jackson},
  {Jogler}, {J{\'o}hannesson}, {Johnson}, {Kamae}, {Kataoka}, {Katsuta},
  {Kn{\"o}dlseder}, {Kuss}, {Lande}, {Larsson}, {Latronico}, {Lemoine-Goumard},
  {Longo}, {Loparco}, {Lovellette}, {Lubrano}, {Madejski}, {Massaro}, {Mayer},
  {Mazziotta}, {McEnery}, {Mehault}, {Michelson}, {Mignani}, {Mitthumsiri},
  {Mizuno}, {Moiseev}, {Monzani}, {Morselli}, {Moskalenko}, {Murgia},
  {Nakamori}, {Nemmen}, {Nuss}, {Ohno}, {Ohsugi}, {Omodei}, {Orienti},
  {Orlando}, {Ormes}, {Paneque}, {Perkins}, {Pesce-Rollins}, {Piron}, {Pivato},
  {Rain{\`o}}, {Rando}, {Razzano}, {Razzaque}, {Reimer}, {Reimer}, {Ritz},
  {Romoli}, {S{\'a}nchez-Conde}, {Schulz}, {Sgr{\`o}}, {Simeon}, {Siskind},
  {Smith}, {Spandre}, {Spinelli}, {Stecker}, {Strong}, {Suson}, {Tajima},
  {Takahashi}, {Takahashi}, {Tanaka}, {Thayer}, {Thayer}, {Thompson},
  {Thorsett}, {Tibaldo}, {Tibolla}, {Tinivella}, {Troja}, {Uchiyama}, {Usher},
  {Vandenbroucke}, {Vasileiou}, {Vianello}, {Vitale}, {Waite}, {Werner},
  {Winer}, {Wood}, {Wood}, {Yamazaki}, {Yang}, \& {Zimmer}}]{Ackermann:2013}
{Ackermann}, M., {Ajello}, M., {Allafort}, A., {et~al.} 2013, Science, 339, 807

\bibitem[{{Ackermann} {et~al.}(2016){Ackermann}, {Ajello}, {Atwood}, {Baldini},
  {Ballet}, {Barbiellini}, {Bastieri}, {Becerra Gonzalez}, {Bellazzini},
  {Bissaldi}, {Blandford}, {Bloom}, {Bonino}, {Bottacini}, {Brandt}, {Bregeon},
  {Bruel}, {Buehler}, {Buson}, {Caliandro}, {Cameron}, {Caputo}, {Caragiulo},
  {Caraveo}, {Cavazzuti}, {Cecchi}, {Charles}, {Chekhtman}, {Cheung}, {Chiang},
  {Chiaro}, {Ciprini}, {Cohen}, {Cohen-Tanugi}, {Cominsky}, {Conrad}, {Cuoco},
  {Cutini}, {D'Ammando}, {de Angelis}, {de Palma}, {Desiante}, {Di Mauro}, {Di
  Venere}, {Dom{\'{\i}}nguez}, {Drell}, {Favuzzi}, {Fegan}, {Ferrara}, {Focke},
  {Fortin}, {Franckowiak}, {Fukazawa}, {Funk}, {Furniss}, {Fusco}, {Gargano},
  {Gasparrini}, {Giglietto}, {Giommi}, {Giordano}, {Giroletti}, {Glanzman},
  {Godfrey}, {Grenier}, {Grondin}, {Guillemot}, {Guiriec}, {Harding}, {Hays},
  {Hewitt}, {Hill}, {Horan}, {Iafrate}, {Hartmann}, {Jogler},
  {J{\'o}hannesson}, {Johnson}, {Kamae}, {Kataoka}, {Kn{\"o}dlseder}, {Kuss},
  {La Mura}, {Larsson}, {Latronico}, {Lemoine-Goumard}, {Li}, {Li}, {Longo},
  {Loparco}, {Lott}, {Lovellette}, {Lubrano}, {Madejski}, {Maldera},
  {Manfreda}, {Mayer}, {Mazziotta}, {Michelson}, {Mirabal}, {Mitthumsiri},
  {Mizuno}, {Moiseev}, {Monzani}, {Morselli}, {Moskalenko}, {Murgia}, {Nuss},
  {Ohsugi}, {Omodei}, {Orienti}, {Orlando}, {Ormes}, {Paneque}, {Perkins},
  {Pesce-Rollins}, {Petrosian}, {Piron}, {Pivato}, {Porter}, {Rain{\`o}},
  {Rando}, {Razzano}, {Razzaque}, {Reimer}, {Reimer}, {Reposeur}, {Romani},
  {S{\'a}nchez-Conde}, {Saz Parkinson}, {Schmid}, {Schulz}, {Sgr{\`o}},
  {Siskind}, {Spada}, {Spandre}, {Spinelli}, {Suson}, {Tajima}, {Takahashi},
  {Takahashi}, {Takahashi}, {Thayer}, {Thompson}, {Tibaldo}, {Torres}, {Tosti},
  {Troja}, {Vianello}, {Wood}, {Wood}, {Yassine}, {Zaharijas}, \&
  {Zimmer}}]{2FHL}
{Ackermann}, M., {Ajello}, M., {Atwood}, W.~B., {et~al.} 2016, \apjs, 222, 5

\bibitem[{{Ackermann} {et~al.}(2017){Ackermann}, {Ajello}, {Baldini}, {Ballet},
  {Barbiellini}, {Bastieri}, {Bellazzini}, {Bissaldi}, {Bloom}, {Bonino},
  {Bottacini}, {Brandt}, {Bregeon}, {Bruel}, {Buehler}, {Cameron}, {Caragiulo},
  {Caraveo}, {Castro}, {Cavazzuti}, {Cecchi}, {Charles}, {Chekhtman}, {Cheung},
  {Chiaro}, {Ciprini}, {Cohen}, {Costantin}, {Costanza}, {Cutini}, {D'Ammando},
  {de Palma}, {Desiante}, {Digel}, {Di Lalla}, {Di Mauro}, {Di Venere},
  {Favuzzi}, {Fegan}, {Ferrara}, {Franckowiak}, {Fukazawa}, {Funk}, {Fusco},
  {Gargano}, {Gasparrini}, {Giglietto}, {Giordano}, {Giroletti}, {Green},
  {Grenier}, {Grondin}, {Guillemot}, {Guiriec}, {Harding}, {Hays}, {Hewitt},
  {Horan}, {Hou}, {J{\'o}hannesson}, {Kamae}, {Kuss}, {La Mura}, {Larsson},
  {Lemoine-Goumard}, {Li}, {Longo}, {Loparco}, {Lubrano}, {Magill}, {Maldera},
  {Malyshev}, {Manfreda}, {Mazziotta}, {Michelson}, {Mitthumsiri}, {Mizuno},
  {Monzani}, {Morselli}, {Moskalenko}, {Negro}, {Nuss}, {Ohsugi}, {Omodei},
  {Orienti}, {Orlando}, {Ormes}, {Paliya}, {Paneque}, {Perkins}, {Persic},
  {Pesce-Rollins}, {Petrosian}, {Piron}, {Porter}, {Principe}, {Rain{\`o}},
  {Rando}, {Razzano}, {Razzaque}, {Reimer}, {Reimer}, {Reposeur}, {Sgr{\`o}},
  {Simone}, {Siskind}, {Spada}, {Spandre}, {Spinelli}, {Suson}, {Tak},
  {Thayer}, {Thompson}, {Torres}, {Tosti}, {Troja}, {Vianello}, {Wood}, \&
  {Wood}}]{FGES:2017}
{Ackermann}, M., {Ajello}, M., {Baldini}, L., {et~al.} 2017, \apj, 843, 139

\bibitem[{{Ahnen} {et~al.}(2017){Ahnen}, {Ansoldi}, {Antonelli}, {Arcaro},
  {Babi{\'c}}, {Banerjee}, {Bangale}, {Barres de Almeida}, {Barrio}, {Becerra
  Gonz{\'a}lez}, {Bednarek}, {Bernardini}, {Berti}, {Bhattacharyya},
  {Biasuzzi}, {Biland }, {Blanch}, {Bonnefoy}, {Bonnoli}, {Carosi}, {Carosi},
  {Chatterjee}, {Colak}, {Colin}, {Colombo}, {Contreras}, {Cortina}, {Covino},
  {Cumani}, {Da Vela}, {Dazzi}, {De Angelis}, {De Lotto}, {de O{\~n}a
  Wilhelmi}, {Di Pierro}, {Doert}, {Dom{\'\i}nguez}, {Dominis Prester},
  {Dorner}, {Doro}, {Einecke}, {Eisenacher Glawion}, {Elsaesser},
  {Engelkemeier}, {Fallah Ramazani}, {Fern{\'a}ndez-Barral}, {Fidalgo},
  {Fonseca}, {Font}, {Fruck}, {Galindo}, {Garc{\'\i}a L{\'o}pez},
  {Garczarczyk}, {Gaug}, {Giammaria}, {Godinovi{\'c}}, {Gora}, {Guberman},
  {Hadasch}, {Hahn}, {Hassan}, {Hayashida}, {Herrera}, {Hose}, {Hrupec},
  {Inada}, {Ishio}, {Konno}, {Kubo}, {Kushida}, {Kuve{\v{z}}di{\'c}}, {Lelas},
  {Lindfors}, {Lombardi}, {Longo}, {L{\'o}pez}, {Maggio}, {Majumdar},
  {Makariev}, {Maneva}, {Manganaro}, {Mannheim}, {Maraschi}, {Mariotti},
  {Mart{\'\i}nez}, {Mazin}, {Menzel}, {Minev}, {Mirzoyan}, {Moralejo},
  {Moreno}, {Moretti}, {Neustroev}, {Niedzwiecki}, {Nievas Rosillo}, {Nilsson},
  {Ninci}, {Nishijima}, {Noda}, {Nogu{\'e}s}, {Paiano}, {Palacio}, {Paneque},
  {Paoletti}, {Paredes}, {Pedaletti}, {Peresano}, {Perri}, {Persic}, {Prada
  Moroni}, {Prand ini}, {Puljak}, {Garcia}, {Reichardt}, {Rhode}, {Rib{\'o}},
  {Rico}, {Righi}, {Saito}, {Satalecka}, {Schroeder}, {Schweizer}, {Shore},
  {Sitarek}, {{\v{S}}nidari{\'c}}, {Sobczynska}, {Stamerra}, {Strzys},
  {Suri{\'c}}, {Takalo}, {Tavecchio}, {Temnikov}, {Terzi{\'c}}, {Tescaro},
  {Teshima}, {Torres-Alb{\`a}}, {Treves}, {Vanzo}, {Vazquez Acosta}, {Vovk},
  {Ward}, {Will}, \& {Zari{\'c}}}]{CasA:2017}
{Ahnen}, M.~L., {Ansoldi}, S., {Antonelli}, L.~A., {et~al.} 2017, \mnras, 472,
  2956

\bibitem[{{Ajello} {et~al.}(2016){Ajello}, {Baldini}, {Barbiellini},
  {Bastieri}, {Bellazzini}, {Bissaldi}, {Bloom}, {Bonino}, {Bottacini},
  {Brandt}, {Bregeon}, {Bruel}, {Buehler}, {Caliandro}, {Cameron}, {Caragiulo},
  {Cavazzuti}, {Charles}, {Chekhtman}, {Ciprini}, {Cohen-Tanugi}, {Condon},
  {Costanza}, {Cutini}, {D'Ammando}, {de Palma}, {Desiante}, {Di Lalla}, {Di
  Mauro}, {Di Venere}, {Drell}, {Dubner}, {Dumora}, {Duvidovich}, {Favuzzi},
  {Focke}, {Fusco}, {Gargano}, {Gasparrini}, {Giacani}, {Giglietto},
  {Glanzman}, {Green}, {Grenier}, {Guiriec}, {Hays}, {Hewitt}, {Hill}, {Horan},
  {Jogler}, {J{\'o}hannesson}, {Jung-Richardt}, {Kensei}, {Kuss}, {Larsson},
  {Latronico}, {Lemoine-Goumard}, {Li}, {Li}, {Longo}, {Loparco}, {Lovellette},
  {Lubrano}, {Magill}, {Maldera}, {Manfreda}, {Mayer}, {Mazziotta}, {McEnery},
  {Michelson}, {Mitthumsiri}, {Mizuno}, {Monzani}, {Morselli}, {Moskalenko},
  {Negro}, {Nuss}, {Orienti}, {Orlando}, {Ormes}, {Paneque}, {Perkins},
  {Pesce-Rollins}, {Piron}, {Pivato}, {Porter}, {Rain{\`o}}, {Rando},
  {Razzano}, {Reimer}, {Reimer}, {Reposeur}, {Schmid}, {Schulz}, {Sgr{\`o}},
  {Simone}, {Siskind}, {Spada}, {Spandre}, {Spinelli}, {Thayer}, {Tibaldo},
  {Torres}, {Tosti}, {Troja}, {Uchiyama}, {Vianello}, {Vink}, {Wood}, \&
  {Yassine}}]{RCW86:2016}
{Ajello}, M., {Baldini}, L., {Barbiellini}, G., {et~al.} 2016, \apj, 819, 98

\bibitem[{{Akaike}(1974)}]{Akaike}
{Akaike}, H. 1974, IEEE Transactions on Automatic Control, 19, 716

\bibitem[{{Anderson} {et~al.}(2014){Anderson}, {Bania}, {Balser}, {Cunningham},
  {Wenger}, {Johnstone}, \& {Armentrout}}]{Wise:2014}
{Anderson}, L.~D., {Bania}, T.~M., {Balser}, D.~S., {et~al.} 2014, \apjs, 212,
  1

\bibitem[{{Arnaud}(1996)}]{XSPEC:1996}
{Arnaud}, K.~A. 1996, Astronomical Society of the Pacific Conference Series,
  Vol. 101, {XSPEC: The First Ten Years}, ed. G.~H. {Jacoby} \& J.~{Barnes}, 17

\bibitem[{{Atwood} {et~al.}(2009){Atwood}, {Abdo}, {Ackermann}, {Althouse},
  {Anderson}, {Axelsson}, {Baldini}, {Ballet}, {Band}, {Barbiellini}, \&
  et~al.}]{LAT:Atwood:2009}
{Atwood}, W.~B., {Abdo}, A.~A., {Ackermann}, M., {et~al.} 2009, \apj, 697, 1071

\bibitem[{{Barr} {et~al.}(2013){Barr}, {Guillemot}, {Champion}, {Kramer},
  {Eatough}, {Lee}, {Verbiest}, {Bassa}, {Camilo}, {{\c{C}}elik}, {Cognard},
  {Ferrara}, {Freire}, {Janssen}, {Johnston}, {Keith}, {Lyne}, {Michelson},
  {Parkinson}, {Ransom}, {Ray}, {Stappers}, \& {Wood}}]{Barr:2013}
{Barr}, E.~D., {Guillemot}, L., {Champion}, D.~J., {et~al.} 2013, \mnras, 429,
  1633

\bibitem[{{Bell}(1978)}]{Bell:1978}
{Bell}, A.~R. 1978, \mnras, 182, 147

\bibitem[{Castro {et~al.}(2011)Castro, Slane, Patnaude, \&
  Ellison}]{Castro:2011}
Castro, D., Slane, P., Patnaude, D.~J., \& Ellison, D.~C. 2011, The
  Astrophysical Journal, 734, 85

\bibitem[{{Chatterjee} {et~al.}(2005){Chatterjee}, {Vlemmings}, {Brisken},
  {Lazio}, {Cordes}, {Goss}, {Thorsett}, {Fomalont}, {Lyne}, \&
  {Kramer}}]{Chatterjee:2005}
{Chatterjee}, S., {Vlemmings}, W.~H.~T., {Brisken}, W.~F., {et~al.} 2005,
  \apjl, 630, L61

\bibitem[{{Dame} {et~al.}(2001){Dame}, {Hartmann}, \& {Thaddeus}}]{Dame:2001}
{Dame}, T.~M., {Hartmann}, D., \& {Thaddeus}, P. 2001, \apj, 547, 792

\bibitem[{{Gao} \& {Han}(2014)}]{Gao:2014}
{Gao}, X.~Y. \& {Han}, J.~L. 2014, \aap, 567, A59

\bibitem[{{Gerbrandt} {et~al.}(2014){Gerbrandt}, {Foster}, {Kothes},
  {Geisb{\"u}sch}, \& {Tung}}]{Gerbrandt:2014}
{Gerbrandt}, S., {Foster}, T.~J., {Kothes}, R., {Geisb{\"u}sch}, J., \& {Tung},
  A. 2014, \aap, 566, A76

\bibitem[{{Green}(2017)}]{Green:2017}
{Green}, D.~A. 2017, VizieR Online Data Catalog, VII/278

\bibitem[{{H.~E.~S.~S. Collaboration} {et~al.}(2018{\natexlab{a}}){H.~E.~S.~S.
  Collaboration}, {Abdalla}, {Abramowski}, {Aharonian}, {Ait Benkhali},
  {Akhperjanian}, {Andersson}, {Ang{\"u}ner}, {Arrieta}, {Aubert}, {Backes},
  {Balzer}, {Barnard}, {Becherini}, {Becker Tjus}, {Berge}, {Bernhard},
  {Bernl{\"o}hr}, {Blackwell}, {B{\"o}ttcher}, {Boisson}, {Bolmont}, {Bordas},
  {Bregeon}, {Brun}, {Brun}, {Bryan}, {Bulik}, {Capasso}, {Carr}, {Casanova},
  {Cerruti}, {Chakraborty}, {Chalme-Calvet}, {Chaves}, {Chen}, {Chevalier},
  {Chr{\'e}tien}, {Colafrancesco}, {Cologna}, {Condon}, {Conrad}, {Cui},
  {Davids}, {Decock}, {Degrange}, {Deil}, {Devin}, {deWilt}, {Dirson},
  {Djannati-Ata{\"\i}}, {Domainko}, {Donath}, {Drury}, {Dubus}, {Dutson},
  {Dyks}, {Edwards}, {Egberts}, {Eger}, {Ernenwein}, {Eschbach}, {Farnier},
  {Fegan}, {Fernand es}, {Fiasson}, {Fontaine}, {F{\"o}rster}, {Fukuyama},
  {Funk}, {F{\"u}{\ss}ling}, {Gabici}, {Gajdus}, {Gallant}, {Garrigoux},
  {Giavitto}, {Giebels}, {Glicenstein}, {Gottschall}, {Goyal}, {Grondin},
  {Hadasch}, {Hahn}, {Haupt}, {Hawkes}, {Heinzelmann}, {Henri}, {Hermann},
  {Hervet}, {Hinton}, {Hofmann}, {Hoischen}, {Holler}, {Horns}, {Ivascenko},
  {Jacholkowska}, {Jamrozy}, {Janiak}, {Jankowsky}, {Jankowsky}, {Jingo},
  {Jogler}, {Jouvin}, {Jung-Richardt}, {Kastendieck}, {Katarzy{\'n}ski},
  {Katz}, {Kerszberg}, {Kh{\'e}lifi}, {Kieffer}, {King}, {Klepser}, {Klochkov},
  {Klu{\'z}niak}, {Kolitzus}, {Komin}, {Kosack}, {Krakau}, {Kraus}, {Krayzel},
  {Kr{\"u}ger}, {Laffon}, {Lamanna}, {Lau}, {Lees}, {Lefaucheur}, {Lefranc},
  {Lemi{\`e}re}, {Lemoine-Goumard}, {Lenain}, {Leser}, {Lohse}, {Lorentz},
  {Liu}, {L{\'o}pez-Coto}, {Lypova}, {Marandon}, {Marcowith}, {Mariaud},
  {Marx}, {Maurin}, {Maxted}, {Mayer}, {Meintjes}, {Meyer}, {Mitchell},
  {Moderski}, {Mohamed}, {Mohrmann}, {Mor{\r{a}}}, {Moulin}, {Murach}, {de
  Naurois}, {Niederwanger}, {Niemiec}, {Oakes}, {O'Brien}, {Odaka}, {{\"O}ttl},
  {Ohm}, {Ostrowski}, {Oya}, {Padovani}, {Panter}, {Parsons}, {Pekeur},
  {Pelletier}, {Perennes}, {Petrucci}, {Peyaud}, {Piel}, {Pita}, {Poon},
  {Prokhorov}, {Prokoph}, {P{\"u}hlhofer}, {Punch}, {Quirrenbach}, {Raab},
  {Reimer}, {Reimer}, {Renaud}, {de los Reyes}, {Rieger}, {Romoli},
  {Rosier-Lees}, {Rowell}, {Rudak}, {Rulten}, {Sahakian}, {Salek}, {Sanchez},
  {Santangelo}, {Sasaki}, {Schlickeiser}, {Sch{\"u}ssler}, {Schulz},
  {Schwanke}, {Schwemmer}, {Settimo}, {Seyffert}, {Shafi}, {Shilon}, {Simoni},
  {Sol}, {Spanier}, {Spengler}, {Spies}, {Stawarz}, {Steenkamp}, {Stegmann},
  {Stinzing}, {Stycz}, {Sushch}, {Takahashi}, {Tavernet}, {Tavernier},
  {Taylor}, {Terrier}, {Tibaldo}, {Tiziani}, {Tluczykont}, {Trichard}, {Tuffs},
  {Uchiyama}, {van der Walt}, {van Eldik}, {van Rensburg}, {van Soelen},
  {Vasileiadis}, {Veh}, {Venter}, {Viana}, {Vincent}, {Vink}, {Voisin},
  {V{\"o}lk}, {Volpe}, {Vuillaume}, {Wadiasingh}, {Wagner}, {Wagner}, {Wagner},
  {White}, {Wierzcholska}, {Willmann}, {W{\"o}rnlein}, {Wouters}, {Yang},
  {Zabalza}, {Zaborov}, {Zacharias}, {Zdziarski}, {Zech}, {Zefi}, {Ziegler}, \&
  {{\.Z}ywucka}}]{RXJ:2018}
{H.~E.~S.~S. Collaboration}, {Abdalla}, H., {Abramowski}, A., {et~al.}
  2018{\natexlab{a}}, \aap, 612, A6

\bibitem[{{H.~E.~S.~S. Collaboration} {et~al.}(2018{\natexlab{b}}){H.~E.~S.~S.
  Collaboration}, {Abdalla}, {Abramowski}, {Aharonian}, {Ait Benkhali},
  {Akhperjanian}, {Ang{\"u}ner}, {Arakawa}, {Arrieta}, {Aubert}, {Backes},
  {Balzer}, {Barnard}, {Becherini}, {Becker Tjus}, {Berge}, {Bernhard},
  {Bernl{\"o}hr}, {Blackwell}, {B{\"o}ttcher}, {Boisson}, {Bolmont}, {Bordas},
  {Bregeon}, {Brun}, {Brun}, {Bryan}, {B{\"u}chele}, {Bulik}, {Capasso},
  {Carr}, {Casanova}, {Cerruti}, {Chakraborty}, {Chalme-Calvet}, {Chaves},
  {Chen}, {Chevalier}, {Chr{\'e}tien}, {Coffaro}, {Colafrancesco}, {Cologna},
  {Condon}, {Conrad}, {Cui}, {Davids}, {Decock}, {Degrange}, {Deil}, {Devin},
  {deWilt}, {Dirson}, {Djannati-Ata{\"\i}}, {Domainko}, {Donath}, {Drury},
  {Dutson}, {Dyks}, {Edwards}, {Egberts}, {Eger}, {Ernenwein}, {Eschbach},
  {Farnier}, {Fegan}, {Fernand es}, {Fiasson}, {Fontaine}, {F{\"o}rster},
  {Funk}, {F{\"u}{\ss}ling}, {Gabici}, {Gajdus}, {Gallant}, {Garrigoux},
  {Giavitto}, {Giebels}, {Glicenstein}, {Gottschall}, {Goyal}, {Grondin},
  {Hahn}, {Haupt}, {Hawkes}, {Heinzelmann}, {Henri}, {Hermann}, {Hervet},
  {Hinton}, {Hofmann}, {Hoischen}, {Holler}, {Horns}, {Ivascenko}, {Iwasaki},
  {Jacholkowska}, {Jamrozy}, {Janiak}, {Jankowsky}, {Jankowsky}, {Jingo},
  {Jogler}, {Jouvin}, {Jung-Richardt}, {Kastendieck}, {Katarzy{\'n}ski},
  {Katsuragawa}, {Katz}, {Kerszberg}, {Khangulyan}, {Kh{\'e}lifi}, {Kieffer},
  {King}, {Klepser}, {Klochkov}, {Klu{\'z}niak}, {Kolitzus}, {Komin}, {Krakau},
  {Kraus}, {Kr{\"u}ger}, {Laffon}, {Lamanna}, {Lau}, {Lees}, {Lefaucheur},
  {Lefranc}, {Lemi{\`e}re}, {Lemoine-Goumard}, {Lenain}, {Leser}, {Lohse},
  {Lorentz}, {Liu}, {L{\'o}pez-Coto}, {Lypova}, {Marandon}, {Marcowith},
  {Mariaud}, {Marx}, {Maurin}, {Maxted}, {Mayer}, {Meintjes}, {Meyer},
  {Mitchell}, {Moderski}, {Mohamed}, {Mohrmann}, {Mor{\r{a}}}, {Moulin},
  {Murach}, {Nakashima}, {de Naurois}, {Niederwanger}, {Niemiec}, {Oakes},
  {O'Brien}, {Odaka}, {{\"O}ttl}, {Ohm}, {Ostrowski}, {Oya}, {Padovani},
  {Panter}, {Parsons}, {Paz Arribas}, {Pekeur}, {Pelletier}, {Perennes},
  {Petrucci}, {Peyaud}, {Piel}, {Pita}, {Poon}, {Prokhorov}, {Prokoph},
  {P{\"u}hlhofer}, {Punch}, {Quirrenbach}, {Raab}, {Reimer}, {Reimer},
  {Renaud}, {de los Reyes}, {Richter}, {Rieger}, {Romoli}, {Rowell}, {Rudak},
  {Rulten}, {Sahakian}, {Saito}, {Salek}, {Sanchez}, {Santangelo}, {Sasaki},
  {Schlickeiser}, {Sch{\"u}ssler}, {Schulz}, {Schwanke}, {Schwemmer},
  {Seglar-Arroyo}, {Settimo}, {Seyffert}, {Shafi}, {Shilon}, {Simoni}, {Sol},
  {Spanier}, {Spengler}, {Spies}, {Stawarz}, {Steenkamp}, {Stegmann}, {Stycz},
  {Sushch}, {Takahashi}, {Tavernet}, {Tavernier}, {Taylor}, {Terrier},
  {Tibaldo}, {Tiziani}, {Tluczykont}, {Trichard}, {Tsuji}, {Tuffs}, {Uchiyama},
  {van der Walt}, {van Eldik}, {van Rensburg}, {van Soelen}, {Vasileiadis},
  {Veh}, {Venter}, {Viana}, {Vincent}, {Vink}, {Voisin}, {V{\"o}lk},
  {Vuillaume}, {Wadiasingh}, {Wagner}, {Wagner}, {Wagner}, {White},
  {Wierzcholska}, {Willmann}, {W{\"o}rnlein}, {Wouters}, {Yang}, {Zabalza},
  {Zaborov}, {Zacharias}, {Zanin}, {Zdziarski}, {Zech}, {Zefi}, {Ziegler}, \&
  {{\.Z}ywucka}}]{VelaJr:2018}
{H.~E.~S.~S. Collaboration}, {Abdalla}, H., {Abramowski}, A., {et~al.}
  2018{\natexlab{b}}, \aap, 612, A7

\bibitem[{{Kalberla} {et~al.}(2005){Kalberla}, {Burton}, {Hartmann}, {Arnal},
  {Bajaja}, {Morras}, \& {P{\"o}ppel}}]{HI:2005}
{Kalberla}, P.~M.~W., {Burton}, W.~B., {Hartmann}, D., {et~al.} 2005, \aap,
  440, 775

\bibitem[{{Lyne} \& {Lorimer}(1995)}]{Lyne:1995}
{Lyne}, A.~G. \& {Lorimer}, D.~R. 1995, Journal of Astrophysics and Astronomy,
  16, 97

\bibitem[{{Parizot} {et~al.}(2006){Parizot}, {Marcowith}, {Ballet}, \&
  {Gallant}}]{Parizot:2006}
{Parizot}, E., {Marcowith}, A., {Ballet}, J., \& {Gallant}, Y.~A. 2006, \aap,
  453, 387

\bibitem[{{Porter} {et~al.}(2008){Porter}, {Moskalenko}, {Strong}, {Orlando},
  \& {Bouchet}}]{Porter:2008}
{Porter}, T.~A., {Moskalenko}, I.~V., {Strong}, A.~W., {Orlando}, E., \&
  {Bouchet}, L. 2008, \apj, 682, 400

\bibitem[{Rakowski {et~al.}(2008)Rakowski, Laming, \&
  Ghavamian}]{Rakowski:2008}
Rakowski, C.~E., Laming, J.~M., \& Ghavamian, P. 2008, The Astrophysical
  Journal, 684, 348

\bibitem[{{Rice} {et~al.}(2016){Rice}, {Goodman}, {Bergin}, {Beaumont}, \&
  {Dame}}]{Rice:2016}
{Rice}, T.~S., {Goodman}, A.~A., {Bergin}, E.~A., {Beaumont}, C., \& {Dame},
  T.~M. 2016, \apj, 822, 52

\bibitem[{{The Fermi-LAT collaboration}(2019)}]{4FGL:2019}
{The Fermi-LAT collaboration}. 2019, arXiv e-prints, arXiv:1902.10045

\bibitem[{{Thompson} {et~al.}(2012){Thompson}, {Baldini}, \&
  {Uchiyama}}]{Thompson:2012}
{Thompson}, D.~J., {Baldini}, L., \& {Uchiyama}, Y. 2012, Astroparticle
  Physics, 39, 22

\bibitem[{Vink(2011)}]{Vink:2011}
Vink, J. 2011, The Astronomy and Astrophysics Review, 20, 49

\bibitem[{{Voges} {et~al.}(1999){Voges}, {Aschenbach}, {Boller}, {Braeuninger},
  {Briel}, {Burkert}, {Dennerl}, {Englhauser}, {Gruber}, {Haberl}, {Hartner},
  {Hasinger}, {Kuerster}, {Pfeffermann}, {Pietsch}, {Predehl}, {Rosso},
  {Schmitt}, {Truemper}, \& {Zimmermann}}]{Voges:1999}
{Voges}, W., {Aschenbach}, B., {Boller}, T., {et~al.} 1999, VizieR Online Data
  Catalog, 9010

\bibitem[{Wilms {et~al.}(2000)Wilms, Allen, \& McCray}]{Wilms:2000}
Wilms, J., Allen, A., \& McCray, R. 2000, The Astrophysical Journal, 542, 914

\bibitem[{{Wood} {et~al.}(2017){Wood}, {Caputo}, {Charles}, {Di Mauro},
  {Magill}, {Perkins}, \& {Fermi-LAT Collaboration}}]{Wood:2017}
{Wood}, M., {Caputo}, R., {Charles}, E., {et~al.} 2017, International Cosmic
  Ray Conference, 301, 824

\bibitem[{{Wootten}(1981)}]{Wootten:1981}
{Wootten}, A. 1981, \apj, 245, 105

\bibitem[{{Yuan} {et~al.}(2013){Yuan}, {Funk}, {J{\'o}hannesson}, {Lande},
  {Tibaldo}, \& {Uchiyama}}]{Yuan:2013}
{Yuan}, Y., {Funk}, S., {J{\'o}hannesson}, G., {et~al.} 2013, \apj, 779, 117

\bibitem[{{Zabalza}(2015)}]{Zabalza:2015}
{Zabalza}, V. 2015, in 34th International Cosmic Ray Conference (ICRC2015),
  Vol.~34, 922

\end{thebibliography}

\end{document}